\documentclass[preprint,aps,12pt,showpacs,nofootinbib,tightenlines]{revtex4-1}
\usepackage{amsmath}
\usepackage{amssymb}
\usepackage{epsfig}
\usepackage{graphicx}
\begin{document}
\preprint{JSNU-PHY-TH-2012-1}

\newcommand{\beq}{\begin{eqnarray}}
\newcommand{\eeq}{\end{eqnarray}}
\newcommand{\non}{\nonumber\\ }
\newcommand{\acp}{{\cal A}_{CP}}
\newcommand{\ov}{\overline}

\newcommand{\psl}{ P \hspace{-2.8truemm}/ }
\newcommand{\nsl}{ n \hspace{-2.2truemm}/ }
\newcommand{\vsl}{ v \hspace{-2.2truemm}/ }
\newcommand{\epsl}{\epsilon \hspace{-1.8truemm}/\,  }

\def\lsim{ {\ \lower-1.2pt\vbox{\hbox{\rlap{$<$}\lower6pt\vbox{\hbox{$\sim$}
}}}\ } }
\def\gsim{ {\ \lower-1.2pt\vbox{\hbox{\rlap{$>$}\lower6pt\vbox{\hbox{$\sim$}
}}}\ } }

\def \epjc{ Eur. Phys. J. C }
\def \jpg{  J. Phys. G }
\def \npb{  Nucl. Phys. B }
\def \npps { Nucl. Phys. B (Proc. Suppl.) }
\def \plb{  Phys. Lett. B }
\def \pr{  Phys. Rep. }
\def \prd{  Phys. Rev. D }
\def \prl{  Phys. Rev. Lett.  }
\def \zpc{  Z. Phys. C  }
\def \jhep{ J. High Energy Phys.  }
\def \ijmpa { Int. J. Mod. Phys. A }
\def \cpc{ Chin. Phys. C }
\def \ctp{ Commun. Theor. Phys. }
\def \rmp{ Rev. Mod. Phys. }
\def \ppnp{ Prog. Part. $\&$ Nucl. Phys. }
\def \arnps{ Ann. Rev. Nucl. Part. Sci. }

\title{ $B \to a_1(1260) a_1(1260)$ and $b_1(1235) b_1(1235)$ decays in the perturbative QCD approach }
\author{
Xin~Liu$^1$\footnote{Electronic address: liuxin.physics@gmail.com}
 and
Zhen-Jun~Xiao$^2$\footnote{Electronic address: xiaozhenjun@njnu.edu.cn}}
\affiliation{\small
$^1$  School of Physics and Electronic Engineering, Jiangsu Normal University,\\ Xuzhou, Jiangsu 221116,
People's Republic of China\\
$^2$  Department of Physics and Institute of Theoretical
Physics, Nanjing Normal University,\\ Nanjing, Jiangsu 210046,
People's Republic of China}
\date{\today}
\bigskip
\begin{abstract}
\bigskip
In this work, we study six tree-dominated $B \to a_1(1260) a_1(1260)$ and $b_1(1235) b_1(1235)$ decays
in the perturbative QCD(pQCD) approach, where $a_1$($b_1$) is a
$^3P_1$($^1P_1$) axial-vector meson. Based on the perturbative calculations and
phenomenological analysis, we find that:
(a) the CP-averaged branching ratio of $B^0 \to a_1^+ a_1^-$ decay in the pQCD approach
is $54.7 \times 10^{-6}$, which agrees well with the current data and
the predictions given in the QCD factorization approach within errors;
(b) the numerical results for the decay rates of other five channels are found
to be in the order of $10^{-6} \sim 10^{-5}$, which could be accessed at B factories
and Large Hadron Collider(LHC) experiments;
(c) other physical observables such as polarization fractions and direct CP-violating asymmetries
are also investigated with the pQCD approach in the present work and the predictions can be
confronted with the relevant experiments in the near future;
(d) the different phenomenologies shown between $B \to a_1 a_1$ and $B \to b_1 b_1$ decays
are expected to be tested by the ongoing LHC and forthcoming Super-B experiments, which
could shed light on the typical QCD dynamics involved in these decay modes,
as well as in $^3P_1$ meson $a_1$ and $^1P_1$ meson $b_1$.

\end{abstract}


\maketitle

\section{Introduction}

Charmless $B$ meson decays to final states involving two axial-vector
mesons $(AA)$ have attracted
attentions in theory and experiments
in the last few years~\cite{Diehl01:b2aa,Calderon07:b2aa,Cheng08:b2aa,Cheng09:bdecays,Aubert09:a1pa1m,Lombardo09:b2aa,Gandini09:b2aa}.
It is expected that through the study of $B \to AA$ decays, the issues related with the internal
structure, such as the
angles between the mixtures of $^3P_1$ and/or $^1P_1$ states~\cite{Yang07:twist,Cheng07:b2ap},
 of the
light axial-vector mesons can receive new understanding. On one hand, $B \to AA$ will open another
window to study their physical properties; The CP asymmetries
of these decays, on the other hand, shall provide another way to measure the Cabibbo-Kobayashi-Maskawa (CKM) angles
$\beta$ and $\alpha$.
Furthermore, analogous to $B \to VV$ decays, being constructed of three polarization states,
the charmless $B$ decays to $AA$ mesons are expected to have rich physics
and provide much more information on the underlying helicity
structure of the decay mechanism through polarization studies~\cite{Cheng08:b2aa}.

Very recently, the measurement on branching ratio(BR) and
fraction of longitudinal polarization($f_L$) for
$B^0 \to a_1(1260)^+ a_1(1260)^-$\cite{Aubert09:a1pa1m,Asner10:hfag} decay has been reported by
BaBar Collaboration,
 \beq
 {\rm BR}(B^0 \to a_1(1260)^+ a_1(1260)^-)_{\rm Exp.}
 &=& 47.3 \pm 12.2\; \times 10^{-6}\;;\label{eq:br-b02a1pa1m} \\
 f_L(B^0 \to a_1(1260)^+ a_1(1260)^-)_{\rm Exp.} &=& 0.31 \pm 0.24\;.\label{eq:fl-b02a1pa1m}
 \eeq
with large uncertainties. However, this measurement will be improved rapidly with the running of
Large Hadron Collider(LHC) experiments, in which the
events of $B$ mesons are expected to be produced more
than those collected in the $B$ factories by about 3 orders per year.

On the theory side, $B \to a_1 a_1$\footnote{Hereafter,
for the sake of simplicity, we will adopt the forms $a_1$
and $b_1$(to be shown below) to denote the meson
$a_1(1260)$ and $b_1(1235)$ in the content, respectively, unless otherwise stated.} decays have been
studied in the literature~\cite{Calderon07:b2aa,Cheng08:b2aa,Cheng09:bdecays},
but the predictions on BRs
of the considered channels are significantly different from each other
by employing the approach with naive factorization~\cite{naive} and
QCD factorization(QCDF)~\cite{qcdf}, respectively.
For $B^0 \to a_1^+ a_1^-$ mode, for example, the branching ratio
predicted in naive factorization is $6.4 \times 10^{-6}$~\cite{Calderon07:b2aa},
while that presented in QCDF is $37.4 \times 10^{-6}$~\cite{Cheng08:b2aa}.
One can easily see that the former result
 is too small to be confronted with the preliminary data and the latter one
 going beyond the naive factorization is large enough to be compatible with the measurements.
As the counterparts of $B \to a_1 a_1$ decays, $B \to b_1 b_1$ modes have also
been investigated within the framework of QCDF~\cite{Cheng08:b2aa} and the BRs are
found to be in the order of $10^{-6} \sim 10^{-5}$ within large uncertainties.
More important,
an interesting pattern of the BRs
for $B \to b_1 b_1$ decays is exhibited through the calculations based on QCDF
in Ref.~\cite{Cheng08:b2aa}, i.e., ${\rm BR}(B^0 \to b_1^0 b_1^0)
> {\rm BR}(B^+ \to b_1^+ b_1^0) > {\rm BR}(B^0 \to b_1^+ b_1^-)$,
which is significantly contrary
to that for $B \to a_1 a_1$ channels, i.e., ${\rm BR}(B^0 \to a_1^0 a_1^0)
< {\rm BR}(B^+ \to a_1^+ a_1^0) < {\rm BR}(B^0 \to a_1^+ a_1^-)$.
Here, we want to mention that, as stated in Ref.~\cite{Cheng08:b2aa},
 the troublesome endpoint singularities from hard spectator scattering
 and annihilation decay amplitudes always
exist in the framework of QCDF
and have to be determined through the input parameters fitted
from the relevant precision measurements.

Inspired by the above interesting facts from both theoretical and
experimental aspects, we here  study $B \to a_1 a_1$ and $b_1 b_1$
decays in the present work by employing the low energy effective
Hamiltonian~\cite{Buras96:weak} and the perturbative QCD(pQCD)
approach~\cite{Keum01:kpi,Li03:ppnp}.

It is worth of stressing that the nonfactorizable spectator and
annihilation diagrams are calculable perturbatively
in the pQCD approach. Furthermore, the new measurements on the pure annihilation
$B_s \to \pi^+ \pi^-$ decay reported by CDF~\cite{Ruffini11:CDF}
and LHCb~\cite{Powell11:LHCb} collaborations last year
confirmed the previous pQCD predictions ~\cite{Li04:pippim,Lu07:bs2mm,Xiao11:pippim},
while the measured large decay rate of $B^0 \to K^+ K^-$ also naturally explained by
the renewed pQCD predictions ~\cite{Xiao11:pippim}.

Although $a_1$ and $b_1$ mesons embrace the same components at the quark level, because of
different couplings of orbital and spin angular momenta,
as explored in the QCD sum rule method~\cite{Yang07:twist},
the hadron dynamics of $b_1$
 is very different from that of its partner, $a_1$.
In our numerical evaluations,
the different phenomenologies do appear between these considered $B \to a_1 a_1$ and
$B \to b_1 b_1$ decays.
Therefore, one could expect reasonably that more new information on the contents such
as polarizations, CP asymmetries, CKM unitary angles, even the knowledge of
color-suppressed processes in $B$ meson
decays~\cite{Chiang:color-suppressed,Charng:color-suppressed,Fleischer:color-suppressed,Li11:puzzle}
may be deduced through the detailed studies on $B \to a_1 a_1$ and $b_1 b_1$ decays. Moreover,
the hadron dynamics could be implicated by these perturbative
calculations with the help of precision measurements.

The paper is organized as follows. In Sec.~\ref{sec:1}, we present
the theoretical framework on the low energy effective Hamiltonian, formalism of pQCD approach and mesons'
wave functions. Then we perform the perturbative calculations for $B \to a_1 a_1$ and $b_1 b_1$ decays in Sec.~\ref{sec:2}.
The analytic expressions of the decay amplitudes for the considered modes are also grouped in this section. The numerical
results and phenomenological analysis are given in
Sec.~\ref{sec:3}. The main conclusions and a short summary are presented in the last section.

\section{Theoretical Framework} \label{sec:1}

For the considered $B \to a_1 a_1$ and $b_1 b_1$ decays with $\bar{b} \to \bar{d}$ transition,
the related weak effective
Hamiltonian $H_{{\rm eff}}$~\cite{Buras96:weak} can be written as
\begin{equation}
H_{\rm eff}\, =\, {G_F\over\sqrt{2}}
\left\{V_{ub}^*V_{ud} \left[C_1(\mu)O_1^{u}(\mu)
+C_2(\mu)O_2^{u}(\mu)\right]- V_{tb}^*V_{td} \sum_{i=3}^{10}C_i(\mu)O_i(\mu)\right\}\;,
\label{eq:heff}
\end{equation}
with the Fermi constant $G_F=1.16639\times 10^{-5}{\rm
GeV}^{-2}$, CKM matrix elements $V$,
and Wilson coefficients $C_i(\mu)$ at the renormalization scale
$\mu$. The local four-quark
operators $O_i(i=1,\cdots,10)$ are
\begin{eqnarray}
{\renewcommand\arraystretch{1.5}
\begin{array}{ll}
\displaystyle
O_1^{u}\, =\,
(\bar{d}_\alpha u_\beta)_{V-A}(\bar{u}_\beta b_\alpha)_{V-A}\;,
& \displaystyle
O_2^{u}\, =\, (\bar{d}_\alpha u_\alpha)_{V-A}(\bar{u}_\beta b_\beta)_{V-A}\;;
\label{eq:operators-1} \\
\displaystyle
O_3\, =\, (\bar{d}_\alpha b_\alpha)_{V-A}\sum_{q'}(\bar{q}'_\beta q'_\beta)_{V-A}\;,
& \displaystyle
O_4\, =\, (\bar{d}_\alpha b_\beta)_{V-A}\sum_{q'}(\bar{q}'_\beta q'_\alpha)_{V-A}\;,
\\
\displaystyle
O_5\, =\, (\bar{d}_\alpha b_\alpha)_{V-A}\sum_{q'}(\bar{q}'_\beta q'_\beta)_{V+A}\;,
& \displaystyle
O_6\, =\, (\bar{d}_\alpha b_\beta)_{V-A}\sum_{q'}(\bar{q}'_\beta q'_\alpha)_{V+A}\;;
\label{eq:operators-2} \\
\displaystyle
O_7\, =\,
\frac{3}{2}(\bar{d}_\alpha b_\alpha)_{V-A}\sum_{q'}e_{q'}(\bar{q}'_\beta q'_\beta)_{V+A}\;,
& \displaystyle
O_8\, =\,
\frac{3}{2}(\bar{d}_\alpha b_\beta)_{V-A}\sum_{q'}e_{q'}(\bar{q}'_\beta q'_\alpha)_{V+A}\;,
\\
\displaystyle
O_9\, =\,
\frac{3}{2}(\bar{d}_\alpha b_\alpha)_{V-A}\sum_{q'}e_{q'}(\bar{q}'_\beta q'_\beta)_{V-A}\;,
& \displaystyle
O_{10}\, =\,
\frac{3}{2}(\bar{d}_\alpha b_\beta)_{V-A}\sum_{q'}e_{q'}(\bar{q}'_\beta q'_\alpha)_{V-A}\;.
\end{array}}
\label{eq:operators-3}
\end{eqnarray}
with the color indices $\alpha, \ \beta$ and the notations
$(\bar{q}'q')_{V\pm A} = \bar q' \gamma_\mu (1\pm \gamma_5)q'$.
The index $q'$ in the summation of the above operators runs
through $u,\;d,\;s$, $c$, and $b$.
Since we work in the leading order[${\cal O}(\alpha_s)$] of the pQCD
approach, it is consistent to use the leading order Wilson coefficients.
For the renormalization group evolution of the Wilson coefficients
from higher scale to lower scale, we use the formulas as given in
Ref.~\cite{Keum01:kpi} directly.

The pQCD approach, one of the method based on
 QCD dynamics in the market, has been
employed to treat two-body nonleptonic $B_{(s)}$ decays extensively.
As a unique feature different from other two factorization approaches,
i.e., QCDF and SCET (soft-collinear effective theory)~\cite{scet}, the pQCD approach
is based on the framework of $k_T$ factorization theorem with taking
the tranverse momentum $k_T$, generally considered as a small and negligible scale,
 of the valence quarks in the
hadrons into account, which results in the Sudakov factor
smearing the endpoint singularities in the decay amplitude and makes the nonfactorizable
 spectator
 and annihilation diagrams perturbatively calculable,
aside from the emission one.

Because of the rather heavy $b$ quark, for convenience, we will work in
 the rest frame of $B$ meson.
Throughout this paper, we will use light-cone
coordinate $(P^+, P^-, {\bf P}_T)$ to describe the meson's momenta with the definitions
\beq
P^{\pm} &=& \frac{p_0 \pm p_3}{\sqrt{2}} \qquad {\rm and} \qquad {\bf P}_T= (p_1, p_2)\;;
 \eeq
 Then for $B^0 \to a_1^+ a_1^-$ decay, for example, the involved three meson
momenta in the light-cone coordinates can be written as
\beq
     P_1=\frac{m_{B}}{\sqrt{2}} (1,1,{\bf 0}_T), \qquad
     P_2 =\frac{m_{B}}{\sqrt{2}} (1-r_3^2,r_2^2,{\bf 0}_T), \qquad
     P_3 =\frac{m_{B}}{\sqrt{2}} (r_3^2,1-r_2^2,{\bf 0}_T),
\eeq
respectively, where
the $a_1^+$ ($a_1^-$) meson moves in the plus (minus) $z$ direction
carrying the momentum $P_2$ ($P_3$) and $r_2=r_3=m_{a_1}/m_{B}$.
The longitudinal and transverse
polarization vectors of axial-vector meson are denoted by $\epsilon^L$ and $\epsilon^T$, respectively, satisfying $P
\cdot \epsilon=0$ in each polarization.
The longitudinal polarization
vectors, $\epsilon_2^L$ and $\epsilon_3^L$, can be chosen as
\beq
\epsilon_2^L &=& \frac{m_{B}}{\sqrt{2} m_{a_1}} (1-r_3^2, -r_2^2,{\bf
0}_T) \qquad  {\rm and} \qquad \epsilon_3^L = \frac{m_{B}}{\sqrt{2} m_{a_1}} (-r_3^2,
1-r_2^2,{\bf 0}_T).
\eeq
And the transverse ones are parameterized
as
$\epsilon_2^T = (0, 0, {\bf 1}_T)$
and
$\epsilon_3^T = (0, 0, {\bf 1}_T)$.

Putting the (light) quark momenta in $B$, $a_1^+$ and $a_1^-$ mesons as $k_1$,
$k_2$, and $k_3$, respectively, we define
\beq
k_1 = (x_1P_1^+,0,{\bf k}_{1T}), \quad k_2 = (x_2 P_2^+,0,{\bf k}_{2T}),
\quad k_3 = (0, x_3 P_3^-,{\bf k}_{3T}).
\eeq
Then, for $B^0 \to a_1^+ a_1^-$ decay, the integration over $k_1^-$, $k_2^-$, and
$k_3^+$ will conceptually lead to the decay amplitude in the pQCD
approach,
\beq
{\cal A}(B \to a_1^+ a_1^-) &\sim &\int\!\! d x_1 d
x_2 d x_3 b_1 d b_1 b_2 d b_2 b_3 d b_3
\non && \cdot \mathrm{Tr}
\left [ C(t) \Phi_{B}(x_1, b_1) \Phi_{a_1^+}(x_2, b_2)
\Phi_{a_1^-}(x_3, b_3) H(x_i, b_i, t) S_t(x_i)\, e^{-S(t)} \right ].
\label{eq:a2}
\eeq
where $b_i$ is the conjugate space coordinate
of $k_{iT}$, and $t$ is the largest energy scale in function
$H(x_i,b_i,t)$. The large logarithms $\ln (m_W/t)$ are included in
the Wilson coefficients $C(t)$. The large double logarithms
($\ln^2 x_i$) given rise from loop corrections to the weak decay
vertex are summed by the threshold
resummation~\cite{Li02:resum}, and they lead to $S_t(x_i)$ which can
decrease faster than any power of $x$ as $x \to 0$, then
remove the endpoint singularities. The last term,
$e^{-S(t)}$, is the Sudakov factor which suppresses the soft
dynamics effectively~\cite{Li98:soft}. Thus it makes the
perturbative calculation of the hard part $H$ applicable at
intermediate scale, i.e., $m_B$ scale. We will calculate
analytically the function $H(x_i,b_i,t)$ for the considered decays
at leading order in $\alpha_s$ expansion and give the convoluted
amplitudes in next section.

The pQCD predictions depend on the inputs for the nonperturbative parameters
such as the decay constants and distribution amplitudes. For heavy $B$ meson,
in principle, both Lorentz structures of the wave function should be considered
in the calculations. However, the contribution induced by the second Lorentz structure
is numerically small and approximately negligible~\cite{Lu03:structure},
we therefore employ the following set of heavy $B$ meson wave function~\cite{Keum01:kpi},
 \beq
 \Phi_{B}(x, b) &=& \frac{i}{\sqrt{6}} \biggl[ (\psl + m_{B})
 \gamma_5 \phi_{B} (x, b) \biggr]_{\alpha\beta} \;,
 \eeq
where the distribution amplitude $\phi_B(x, b)$ has
been modeled as~\cite{Keum01:kpi},
   \beq
\phi_{B}(x, b)&=& N_B x^2(1-x)^2
\exp\left[-\frac{1}{2}\left(\frac{xm_B}{\omega_B}\right)^2
-\frac{\omega_B^2 b^2}{2}\right] \;, \label{eq:DA-B}
\eeq
In recent years, the shape parameter $\omega_B$ in Eq.~(\ref{eq:DA-B}) has been
fixed at $0.40$~GeV in the pQCD approach by using the rich experimental
data on the $B$ mesons with $f_{B}= 0.19$~GeV. The normalization factor $N_{B}$
is related to the decay constant $f_{B}$ through
\beq
\int_0^1 dx \phi_{B}(x, b=0) &=& \frac{f_{B}}{2 \sqrt{6}}\;.
\eeq
Correspondingly, the normalization constant $N_B$ is $91.745$ for $\omega_B=0.40$.
To analyze the uncertainties of theoretical predictions induced
by the inputs, we will vary the shape parameter $\omega_{B}$ by $10\%$.

For the wave functions of axial-vector $a_1$ and $b_1$ mesons, one
longitudinal($L$) and two transverse($T$) polarizations are
involved, and can be written as~\cite{Li09:afm},
 \beq
\Phi^L_A(x)&=& \frac{1}{\sqrt{6}}\gamma_5 \biggl\{ m_A
\epsl_A^{*L} \phi_A(x) + \epsl^{*L}_A \psl \phi_A^t(x)+ m_A
\phi_A^s(x)\biggr\}_{\alpha\beta} \;, \label{eq:wfl-a} \\
\Phi^T_A(x)&=&\frac{1}{\sqrt{6}} \gamma_5\biggl\{ m_A
\epsl_A^{*T} \phi_A^v(x) + \epsl^{*T}_A \psl \phi_A^T(x)+ m_A i
\epsilon_{\mu\nu\rho\sigma} \gamma_5 \gamma^\mu \epsilon_T^{*\nu}
n^\rho v^\sigma \phi_A^a(x)\biggr\}_{\alpha\beta} \;, \label{eq:wft-a}
 \eeq
where $x$ denotes the momentum
fraction carried by quark in the meson, and $n=(1, 0, {\bf 0}_T)$
and $v=(0, 1, {\bf 0}_T)$ are dimensionless light-like unit vectors.
We here adopt the convention $\epsilon^{0123}=1$ for the
Levi-Civita tensor $\epsilon^{\mu\nu\alpha\beta}$.

The twist-2 distribution amplitudes for the longitudinally and
trasversely polarized axial-vector $^3P_1$ and $^1P_1$ mesons can
be parameterized as~\cite{Yang07:twist,Li09:afm}: \beq
 \phi_A(x) & = & \frac{3 f_A}{ \sqrt{2 N_c}}  x (1- x) \left[ a_{0A}^\parallel + 3
a_{1A}^\parallel\, (2x-1) +
a_{2A}^\parallel\, \frac{3}{2} ( 5(2x-1)^2  - 1 ) \right] ,\label{eq:ldaa}\\
 \phi_A^T(x) & = & \frac{3 f_A}{ \sqrt{2 N_c}}  x (1- x)
 \left[ a_{0A}^\perp + 3 a_{1A}^\perp\, (2x-1) +
a_{2A}^\perp\, \frac{3}{2} ( 5(2x-1)^2  - 1 ) \right],
\label{eq:tdaa} \eeq
 where $f_A$ is the decay constant. Here, the definition of these distribution
amplitudes $\phi_A(x)$ and $\phi_A^T(x)$ satisfy the
 following relations:
 \beq
\int_0^1 \phi_{^3P_1}(x) &=& \frac{f_{^3P_1}}{2 \sqrt{2 N_c}},
\;\;\;\;\;\;\;\;\;\;\;\;\;\;\;\; \int_0^1 \phi^T_{^3P_1}(x) =
a^{\perp}_{0 ^3P_1}\frac{f_{^3P_1}}{2 \sqrt{2 N_c}}\;;\non
\int_0^1 \phi_{^1P_1}(x) &=& a^{||}_{0 ^1P_1} \frac{f_{^1P_1}}{2
\sqrt{2 N_c}}, \;\;\;\;\;\;\;\; \int_0^1 \phi^T_{^1P_1}(x) =
\frac{f_{^1P_1}}{2 \sqrt{2 N_c}}\;.
 \eeq
where $a^{||}_{0 ^3P_1}= 1$ and $a^{\perp}_{0 ^1P_1}= 1$ have been used.

As for twist-3 distribution amplitudes for axial-vector meson, we
use the following forms~\cite{Li09:afm}:
 \beq
\phi_{A}^t(x) &=
&\frac{3 f_A}{2\sqrt{2N_c}}\left\{ a_{0A}^\perp (2x-1)^2+
\frac{1}{2}\,a_{1A}^\perp\,(2x-1) (3 (2x-1)^2-1) \right\}
 ,\\
\phi_{A}^s(x)&=& \frac{3 f_A}{2\sqrt{2N_c}} \frac{d}{dx}\left\{ x
(1- x) ( a_{0A}^\perp + a_{1A}^\perp (2x-1) ) \right\}\;;
 \eeq
 \beq
 \phi_{A}^v(x)&=&\frac{3 f_A}{4\sqrt{2N_c}} \left\{ \frac{1}{2} a_{0A}^\parallel (1+(2x-1)^2) +  a_{1A}^\parallel (2x-1)^3 \right\}
 , \\
 \phi_{A}^a(x)&=& \frac{3 f_A}{4\sqrt{2N_c}}\frac{d}{dx}  \left\{ x (1- x) ( a_{0A}^\parallel + a_{1A}^\parallel (2x-1))  \right\}\;.
 \eeq

The Gegenbauer moments $a_{i,A}^{||(\perp)}$ have been studied extensively in the
literatures (see Ref.~\cite{Yang07:twist} and references therein),
here we adopt the following values: \beq
 f_{a_1} &=& 0.238 \pm 0.010~~{\rm GeV}\;, \qquad a^{||}_{2,a_1}= -0.02\pm 0.02\;,\;\;\;\;\;\;\; a^{\perp}_{1,a_1}= -1.04\pm 0.34; \non
 f_{b_1} &=& 0.180 \pm 0.008~~{\rm GeV}\;, \qquad a^{||}_{1,b_1}= -1.95\pm 0.35\;,\;\;\;\;\;\;\; a^{\perp}_{2,b_1}=\hspace{0.4cm} 0.03 \pm 0.19. \label{eq:inputs}
\eeq

Note that we have included the intrinsic $b$ dependence for
the heavy meson wave function $\phi_B$ but not for the light axial-vector meson
wave function $\phi_{A}$. It has
been shown that the intrinsic $b$ dependence of the
light meson wave functions
is not important and negligible~\cite{Li95:kteffect,Huang05:kteffect}.
It is reasonable to preliminarily assume that the intrinsic $b$ dependence of the $a_1$ and $b_1$ wave
functions, which are still unknown, is not essential either.

\section{Perturbative Calculations in pQCD approach} \label{sec:2}

\begin{figure}[h,t,b]
\vspace{-0.1cm} \centerline{\epsfxsize=16 cm
\epsffile{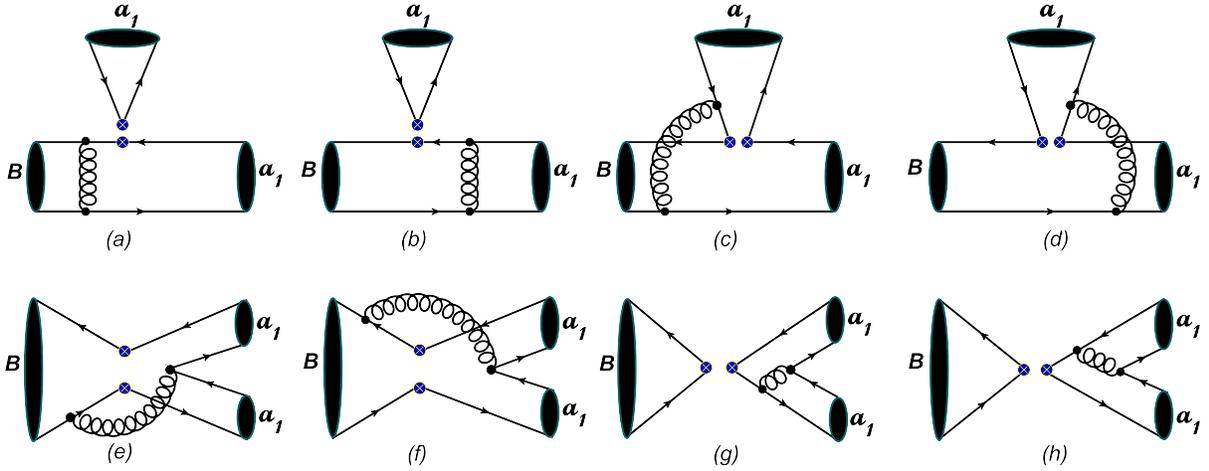}} \vspace{0.2cm} \caption{(Color online)
Typical Feynman diagrams for $B \to a_1 a_1$ decays at the lowest
order in the pQCD approach. By replacing the $^3P_1$ meson $a_1$ in (a)-(h)
with the $^1P_1$ meson $b_1$, one will obtain the corresponding Feynman diagrams
for $B \to b_1 b_1$ decay modes.} \label{fig:fig1}
\end{figure}

There are three kinds of polarizations of a axial-vector
meson, namely, longitudinal ($L$), normal ($N$), and transverse ($T$).
Analogous to the $B \to \rho\rho$
decays~\cite{Li05:b2rhorho,Chen06:b2rhorho,Li06:b2rhorho},
the amplitudes for the $B\to a_1 a_1$ decays
are also characterized by the polarization states of these axial-vector mesons.
In terms of helicities, the decay amplitudes ${\cal M}^{(\sigma)}$ for $B
\to a_1(P_2,\epsilon^*_2) a_1(P_3,\epsilon^*_3)$ decays can be
generally described by
\beq
{\cal M}^{(\sigma)}&=&\epsilon_{2\mu}^{*}(\sigma)\epsilon_{3\nu}^{*}(\sigma)
\left[ a \,\, g^{\mu\nu} + {b \over m_{a_1} m_{a_1}} P_1^\mu P_1^\nu + i{c
\over m_{a_1} m_{a_1} } \epsilon^{\mu\nu\alpha\beta} P_{2\alpha}
P_{3\beta}\right]\;,\non &\equiv &m_{B}^{2}{\cal
M}_{L}+m_{B}^{2}{\cal M}_{N}
\epsilon^{*}_{2}(\sigma=T)\cdot\epsilon^{*}_{3}(\sigma=T) \non &&
+i{\cal M}_{T}\epsilon^{\alpha \beta\gamma \rho}
\epsilon^{*}_{2\alpha}(\sigma)\epsilon^{*}_{3\beta}(\sigma)
P_{2\gamma }P_{3\rho }\; , \label{eq:msigma}
\eeq
where the superscript $\sigma$ denotes the helicity states of two mesons with $L(T)$
standing for the longitudinal (transverse) component. And the
definitions of the amplitudes $ {\cal M}_{i} (i=L,N,T)$ in terms
of the Lorentz-invariant amplitudes $a$, $b$ and $c$ are
\beq
m_{B}^2 \,\, {\cal M}_L &=& a \,\, \epsilon_2^{*}(L) \cdot
\epsilon_3^{*}(L) +{b \over m_{a_1} m_{a_1} } \epsilon_{2}^{*}(L) \cdot
P_3 \,\, \epsilon_{3}^{*}(L) \cdot P_2\;, \non
m_{B}^2 \,\, {\cal M}_N &=& a \;,\label{eq:amp}\\
m_{B}^2 \,\, {\cal M}_T &=& {c \over r_2\,
r_3}\;.\label{id-rel}\nonumber
\eeq
We therefore will evaluate the
helicity amplitudes ${\cal M}_L, {\cal M}_N, {\cal M}_T$ based on
the pQCD
approach, respectively.

As illustrated in Fig.~\ref{fig:fig1},
there are 8 types of diagrams contributing to the $B \to a_1 a_1$ and $b_1 b_1$ decays
at the lowest order in the pQCD approach.
We firstly calculate the usual factorizable spectator($fs$) diagrams (a) and
(b), in which one can factor out the form factors $B \to a_1$ and $B \to b_1$.
The corresponding Feynman
amplitudes with longitudinal polarization($L$) are given as follows,
\begin{enumerate}
\item[]{(i) $(V-A)(V-A)$ operators:}
 \beq
F^{L}_{fs}&=& 8 \pi C_F f_{a_1}
m_{B}^2\int_0^1 d x_{1} dx_{3}\, \int_{0}^{\infty} b_1 db_1 b_3
db_3\,
\non & &
\times \phi_{B}(x_1,b_1)\left\{ \left[(1+x_3 )\phi_{a_1}(x_3) + r_{a_1} (1-2
x_3) (\phi^s_{a_1}(x_3)+\phi^t_{a_1}(x_3))\right]  \right.\non & &
\left. \times h_{fs}(x_1,x_3,b_1,b_3)E_{fs}(t_a)+ 2 r_{a_1} \phi^s_{a_1} (x_3)\;
h_{fs}(x_3,x_1,b_3,b_1) \;E_{fs}(t_b)
\right\}\;, \label{eq:b-fs-l}
 \eeq
where
$C_F=4/3$ is a color
factor. The convolution functions $E_i$, the factorization hard
scales $t_i$, and the hard functions $h_i$ can be referred to
Ref.~\cite{Xiao:pqcd}.

\item[]{(ii) $(V-A)(V+A)$ operators:}
\beq
F_{fs}^{L;P1}&=& -F^{L}_{fs}\;, \label{eq:b-fs-p1-l}
\eeq
which is originated from $\langle a_1 | V + A | 0 \rangle = -
\langle a_1 | V - A| 0 \rangle $.

\item[]{(iii) $(S-P)(S+P)$ operators:}
 \beq
 F_{fs}^{L;P2}&=& 0 \;;\label{eq:b-fs-p2-l}
\eeq
because the emitted axial-vector meson can not be produced through
a scalar or a pseudoscalar current.
\end{enumerate}

For the nonfactorizable spectator($nfs$) diagrams 1(c) and 1(d), the corresponding
decay amplitudes can be read as
\begin{enumerate}
\item[]{(i) $(V-A)(V-A)$ operators:}
\beq
M^{L}_{nfs}&=&\frac{32}{\sqrt{6}}\pi C_F m_{B}^2 \int_{0}^{1}d x_{1}d x_{2}\,d
x_{3}\,\int_{0}^{\infty} b_1d b_1 b_2d b_2\,
\phi_{B}(x_1,b_1)
\non & & \times
\phi_{a_1}(x_2)\left\{ \left[(1-x_2)\phi_{a_1}(x_3) - r_{a_1} x_3 (\phi^s_{a_1}(x_3)-
\phi^t_{a_1}(x_3))\right]\right.\non & & \left.
\times E_{nfs}(t_c)h_{nfs}^c(x_1,x_2,x_3,b_1,b_2)
-h_{nfs}^d(x_1,x_2,x_3,b_1,b_2)\right.\non & & \left. \times
\left[(x_2+x_3)\phi_{a_1}(x_3) - r_{a_1} x_3 (\phi^s_{a_1}(x_3)+
\phi^t_{a_1}(x_3))\right] E_{nfs}(t_d)
\right\} \label{eq:b-nfs-l}\; ,
\eeq

\item[]{(ii) $(V-A)(V+A)$ operators:}
\beq
M_{nfs}^{L;P1}&=&\frac{32}{\sqrt{6}}\pi C_F m_{B}^2 \int_{0}^{1}d
x_{1}d x_{2}\,d x_{3}\,\int_{0}^{\infty} b_1d b_1 b_2d b_2\,\phi_{B}(x_1,b_1)
 \non && \times
 r_{a_1}  \left\{
\left[(1-x_2) (\phi_{a_1}^s(x_2)
+\phi_{a_1}^t(x_2)) \phi_{a_1}(x_3)- r_{a_1} \left(\phi_{a_1}^s(x_2)
 \right.\right.\right.\non
 & & \left.\left.\left. \times [(x_2-x_3-1)
\phi^s_{a_1}(x_3)-(x_2+x_3-1)\phi^t_{a_1}(x_3)]+\phi_{a_1}^t(x_2)\right.
\right.\right.\non && \left.\left.\left.  \times
[(x_2+x_3-1)\phi_{a_1}^s(x_3)+(1-x_2+x_3)\phi_{a_1}^t]\right)\right]
h_{nfs}^c(x_1,x_2,x_3,b_1,b_2)\right.\non &&\left.
\times E_{nfs}(t_c) -h_{nfs}^d(x_1,x_2,x_3,b_1,b_2)E_{nfs}(t_d)\left[x_2(\phi_{a_1}^s(x_2)-\phi_{a_1}^t(x_2))
\right.\right.\non &&\left.\left. \times
\phi_{a_1}(x_3)+ r_{a_1}
(x_2 (\phi_{a_1}^s(x_2)-\phi_{a_1}^t(x_2))(\phi^s_{a_1}(x_3)-
\phi^t_{a_1}(x_3))\right.\right. \non &&\left.\left.
 + x_3
(\phi_{a_1}^s(x_2)+\phi_{a_1}^t(x_2))(\phi^s_{a_1}(x_3)+
\phi^t_{a_1}(x_3)))\right]
\right\} \label{eq:b-nfs-p1-l}\;,
\eeq

\item[]{(iii) $(S-P)(S+P)$ operators:}
\beq
 M_{nfs}^{L;P2}&=& \frac{32}{\sqrt{6}}\pi C_F m_{B}^2
\int_{0}^{1}d x_{1}d x_{2}\,d x_{3}\,\int_{0}^{\infty} b_1d b_1 b_2d
b_2\, \phi_{B}(x_1,b_1)
\non & &
\times  \phi_{a_1}(x_2)\left\{\left[(x_2-x_3-1)\phi_{a_1}(x_3) +
r_{a_1} x_3 (\phi^s_{a_1}(x_3)+
\phi^t_{a_1}(x_3))\right]\right. \non
& &\left. \times E_{nfs}(t_c)h_{nfs}^c(x_1,x_2,x_3,b_1,b_2)
+h_{nfs}^d(x_1,x_2,x_3,b_1,b_2)
\right. \non
& &\left. \times\left[ x_2 \phi_{a_1}(x_3) - r_{a_1} x_3
(\phi^s_{a_1}(x_3)- \phi^t_{a_1}(x_3))\right]
E_{nfs}(t_d) \right\} \label{eq:b-nfs-p2-l}\;;
\eeq
\end{enumerate}
In the above three formulas, i.e., Eqs.~(\ref{eq:b-nfs-l})-(\ref{eq:b-nfs-p2-l}),
one can find that there exist cancellations between the contributions of the
two diagrams in Fig.~\ref{fig:fig1}(c) and \ref{fig:fig1}(d).

The Feynman diagrams shown in Fig.~\ref{fig:fig1}(e) and \ref{fig:fig1}(f) are
the nonfactorizable annihilation($nfa$) ones, whose contributions are
\begin{enumerate}
\item[]{(i) $(V-A)(V-A)$ operators:}
\beq
M_{nfa}^{L}&=& \frac{32}{\sqrt{6}}\pi C_F m_{B}^2
\int_{0}^{1}d x_{1}d x_{2}\,d x_{3}\,\int_{0}^{\infty} b_1d b_1 b_2d
b_2\,
\non & &
 \times \phi_{B}(x_1,b_1) \left\{
\left[(1-x_3)\phi_{a_1}(x_2)\phi_{a_1}(x_3)+ r_{a_1}  r_{a_1} \left(\phi_{a_1}^s(x_2)
\right.\right.\right.
\non && \left.\left.\left.
\times [(1+x_2-x_3)
\phi^s_{a_1}(x_3) -
(1-x_2-x_3)\phi^t_{a_1}(x_3)]+\phi_{a_1}^t(x_2)\right.\right.\right.
\non && \left.\left.\left.
\times [(1-x_2-x_3)\phi_{a_1}^s(x_3)-(1+x_2-x_3)\phi_{a_1}^t(x_3)]
\right)\right] E_{nfa}(t_e)  \right.\non
& & \left.\times h_{nfa}^e(x_1,x_2,x_3,b_1,b_2)-E_{nfa}(t_f)
h_{nfa}^f(x_1,x_2,x_3,b_1,b_2) \right. \non && \left.
\times\left[x_2 \phi_{a_1}(x_2)\phi_{a_1}(x_3)+ r_{a_1} r_{a_1}
\left(\phi_{a_1}^s(x_2)[(x_2-x_3+3) \phi^s_{a_1}(x_3)\right.\right. \right.\non &&
\left. \left.\left.+(1-x_2-x_3)
\phi^t_{a_1}(x_3)]+\phi_{a_1}^t(x_2)
[(x_2+x_3-1)\phi_{a_1}^s(x_3)\right.\right. \right.\non &&
\left. \left.\left.
+(1-x_2+x_3)\phi_{a_1}^t(x_3)]\right)\right]
 \right\} \label{eq:b-nfa-l}\;,
\eeq

\item[]{(ii) $(V-A)(V+A)$ operators:}
\beq
M_{nfa}^{L;P1}&=& \frac{32}{\sqrt{6}}\pi C_F m_{B}^2 \int_{0}^{1}d
x_{1}d x_{2}\,d x_{3}\,\int_{0}^{\infty} b_1d b_1 b_2d b_2\,
 \non & &
\times \phi_{B}(x_1,b_1)\left\{ \left[ r_{a_1}
x_2(\phi_{a_1}^s(x_2)+\phi_{a_1}^t(x_2)) \phi_{a_1}(x_3) - r_{a_1} (1 -x_3)
\right.\right.
 \non
 & & \left.\left.
\times \phi_{a_1}(x_2)(\phi^s_{a_1}(x_3)- \phi^t_{a_1}(x_3))\right]
 E_{nfa}(t_e)h_{nfa}^e(x_1,x_2,x_3,b_1,b_2)\right. \non & & \left.
 + \left[ r_{a_1}
(2-x_2)(\phi_{a_1}^s(x_2)+\phi_{a_1}^t(x_2))
 \phi_{a_1}(x_3)- r_{a_1} (1+x_3)\right.\right.\non && \left.\left.\times
\phi_{a_1}(x_2)(\phi^s_{a_1}(x_3)-
 \phi^t_{a_1}(x_3))\right]E_{nfa}(t_f)h_{nfa}^f(x_1,x_2,x_3,b_1,b_2)
\right\} \label{eq:b-nfa-p1-l}\;,
\eeq

\item[]{(iii) $(S-P)(S+P)$ operators:}
\beq
M_{nfa}^{L;P2}&=& -\frac{32}{\sqrt{6}}\pi C_F m_{B}^2
\int_{0}^{1}d x_{1}d x_{2}\,d x_{3}\,\int_{0}^{\infty} b_1d b_1 b_2d
b_2\,
\non & &
\times \phi_{B}(x_1,b_1)\left\{
\left[(1-x_3)\phi_{a_1}(x_2)\phi_{a_1}(x_3)+ r_{a_1}  r_{a_1} \left(\phi_{a_1}^s(x_2)
\right.\right.\right.
\non & &
\left.\left.\left. \times [(x_2-x_3+3)
\phi^s_{a_1}(x_3)-
(1-x_2-x_3)\phi^t_{a_1}(x_3)]+\phi_{a_1}^t(x_2)\right.\right.\right.
\non && \left.\left.\left.
\times [(1-x_2-x_3)\phi_{a_1}^s(x_3)+(1-x_2+x_3)\phi_{a_1}^t(x_3)]
\right)\right] E_{nfa}(t_f)  \right.\non
& & \left. \times
h_{nfa}^f(x_1,x_2,x_3,b_1,b_2)- E_{nfa}(t_e)h_{nfa}^e(x_1,x_2,x_3,b_1,b_2)
\right. \non && \left.
 \times
\left[x_2 \phi_{a_1}(x_2)\phi_{a_1}(x_3)+ r_{a_1}   r_{a_1}
\left(\phi_{a_1}^s(x_2)[(1+x_2-x_3) \phi^s_{a_1}(x_3)
\right.\right. \right.\non && \left. \left.\left.
+(1-x_2-x_3)\phi^t_{a_1}(x_3)]+\phi_{a_1}^t(x_2)
[(x_2+x_3-1)\phi_{a_1}^s(x_3)
\right.\right. \right.\non && \left. \left.\left.
-(1+x_2-x_3)\phi_{a_1}^t(x_3)]\right)\right]
 \right\} \label{eq:b-nfa-p2-l}\;;
\eeq
\end{enumerate}

For the last two diagrams in Fig.~\ref{fig:fig1}, i.e., the factorizable annihilation($fa$) diagrams
\ref{fig:fig1}(g) and \ref{fig:fig1}(h), we have
\begin{enumerate}
\item[]{(i) $(V-A)(V-A)$ operators:}
\beq
F_{fa}^{L}&=&  -8 \pi C_F m_{B}^2\int_0^1 d x_{2} dx_{3}\,
\int_{0}^{\infty} b_2 db_2 b_3 db_3\, \left\{ \left[ x_2
\phi_{a_1}(x_2) \phi_{a_1}(x_3) + 2 r_{a_1}  r_{a_1} \right.\right. \non & &
\left.\left.  \times  \left((x_2 +
1)\phi^s_{a_1}(x_2)+(x_2-1)\phi^t_{a_1}(x_2)\right)\phi_{a_1}^s(x_3)\right]
 h_{fa}(x_2,1-x_3,b_2,b_3) \right. \non && \left.\times
E_{fa}(t_g)  - \left[(1-x_3)\phi_{a_1}(x_2) \phi_{a_1}(x_3) - 2 r_{a_1}  r_{a_1}
\phi_{a_1}^s(x_2) \left((x_3-2)\phi^s_{a_1} (x_3)
\right.\right.\right.\non && \left.\left.\left.- x_3
\phi_{a_1}^t(x_3)\right) \right] E_{fa}(t_h)h_{fa}(1-x_3,x_2,b_3,b_2)
\right\}\label{eq:b-fa-l}\;,
\eeq

\item[]{(ii) $(V-A)(V+A)$ operators:}
\beq
 F_{fa}^{L;P1}&=& F_{fa}^{L}\;, \label{eq:b-fa-p1-l}
\eeq

\item[]{(iii) $(S-P)(S+P)$ operators:}
\beq
 F_{fa}^{L;P2}&=& 16 \pi C_F m_{B}^2  \int_0^1 d
x_{2} dx_{3}\, \int_{0}^{\infty} b_2 db_2 b_3 db_3\,\left\{ \left[2
r_{a_1} \phi_{a_1}(x_2) \phi^s_{a_1}(x_3) \right.\right. \non & &
\left.\left. + r_{a_1}
 x_2 (\phi_{a_1}^s(x_2)- \phi_{a_1}^t(x_2))\phi_{a_1}(x_3) \right]
h_{fa}(x_2,1-x_3,b_2,b_3) E_{fa}(t_g)\right. \non && \left.  + \left[2
r_{a_1} \phi_{a_1}^s(x_2)\phi_{a_1}(x_3) + r_{a_1}
(1-x_3) \phi_{a_1}(x_2) (\phi_{a_1}^s(x_3)+\phi_{a_1}^t(x_3)) \right]
 \right. \non && \left. \times
E_{fa}(t_h)h_{fa}(1-x_3,x_2,b_3,b_2) \right\} \label{eq:b-fa-p2-l}\;.
 \eeq
\end{enumerate}
It is interesting to notice that there is a large cancellation in
the $F^L_{fa}$, i.e., Eq.~(\ref{eq:b-fa-l}),
from the factorizable annihilation diagrams \ref{fig:fig1}(g)
and \ref{fig:fig1}(h), which can result in the exact zero contribution
in the SU(3) limit.

We can also present the factorization formulas for the Feynman amplitudes with transverse polarizations,
\beq
F_{fs}^{N}&=& 8 \pi C_F f_{a_1}
m_{B}^2\int_0^1 d x_{1} dx_{3}\, \int_{0}^{\infty} b_1 db_1 b_3
db_3\, \phi_{B}(x_1,b_1)\, r_{a_1}
\non & &
\times \left\{ \left[\phi^T_{a_1}(x_3) + r_{a_1} x_3  (\phi^v_{a_1}(x_3)-\phi^a_{a_1}(x_3)) + 2 r_{a_1} \phi^v_{a_1}(x_3)\right] E_{fs}(t_a) \right.\non & &
\left. \times h_{fs}(x_1,x_3,b_1,b_3)+ r_{a_1} (\phi^a_{a_1} (x_3) + \phi^v_{a_1} (x_3))\;
h_{fs}(x_3,x_1,b_3,b_1) \;E_{fs}(t_b)
\right\}\;, \label{eq:b-fs-n}
 \eeq
 \beq
F_{fs}^{T}&=& 16 \pi C_F f_{a_1}
m_{B}^2\int_0^1 d x_{1} dx_{3}\, \int_{0}^{\infty} b_1 db_1 b_3
db_3\, \phi_{B}(x_1,b_1)\, r_{a_1}
\non & &
\times \left\{ \left[\phi^T_{a_1}(x_3) - r_{a_1} x_3  (\phi^v_{a_1}(x_3)-\phi^a_{a_1}(x_3)) + 2 r_{a_1} \phi^a_{a_1}(x_3)\right] E_{fs}(t_a) \right.\non & &
\left. \times h_{fs}(x_1,x_3,b_1,b_3)+ r_{a_1} (\phi^a_{a_1} (x_3) + \phi^v_{a_1} (x_3))\;
h_{fs}(x_3,x_1,b_3,b_1) \;E_{fs}(t_b)
\right\}\;, \label{eq:b-fs-t}
\eeq
\beq
  F_{fs}^{N;P_1} &=& - F_{fs}^{N}\;,\label{eq:b-fs-p1-n} \\
  F_{fs}^{T;P_1} &=& - F_{fs}^{T}\;; \label{eq:b-fs-p1-t}
\eeq
\beq
  F_{fs}^{N;P_2} &=& 0\;,\label{eq:b-fs-p2-n} \\
  F_{fs}^{T;P_2} &=& 0\;; \label{eq:b-fs-p2-t}
\eeq
\beq
M^{N}_{nfs}&=& \frac{32}{\sqrt{6}}\pi C_F m_{B}^2 \int_{0}^{1}d x_{1}d x_{2}\,d
x_{3}\,\int_{0}^{\infty} b_1d b_1 b_2d b_2\,
\phi_{B}(x_1,b_1)\, r_{a_1}
\non & & \times
\left\{ \left[(1-x_2)(\phi^a_{a_1}(x_2) + \phi^v_{a_1}(x_2))
\phi^T_{a_1}(x_3)\right] h_{nfs}^c(x_1,x_2,x_3,b_1,b_2)
 \right.\non & & \left. \times E_{nfs}(t_c)
+h_{nfs}^d(x_1,x_2,x_3,b_1,b_2)
\left[x_2 (\phi^a_{a_1}(x_2) + \phi^v_{a_1}(x_2))\phi^T_{a_1}(x_3)
 \right. \right. \non && \left. \left. - 2 r_{a_1} (x_2 + x_3 ) (\phi^a_{a_1}(x_2) \phi^a_{a_1}(x_3)+ \phi^v_{a_1}(x_2)
\phi^v_{a_1}(x_3))\right] E_{nfs}(t_d)
\right\} \label{eq:b-nfs-n}\;, \\
M^{T}_{nfs}&=& \frac{64}{\sqrt{6}}\pi C_F m_{B}^2 \int_{0}^{1}d x_{1}d x_{2}\,d
x_{3}\,\int_{0}^{\infty} b_1d b_1 b_2d b_2\,
\phi_{B}(x_1,b_1)\, r_{a_1}
\non & & \times
\left\{ \left[(1-x_2)(\phi^a_{a_1}(x_2) + \phi^v_{a_1}(x_2))
\phi^T_{a_1}(x_3)\right] h_{nfs}^c(x_1,x_2,x_3,b_1,b_2)
 \right.\non & & \left. \times E_{nfs}(t_c)
+h_{nfs}^d(x_1,x_2,x_3,b_1,b_2)
\left[x_2 (\phi^a_{a_1}(x_2) + \phi^v_{a_1}(x_2))\phi^T_{a_1}(x_3)
 \right. \right. \non && \left. \left. - 2 r_{a_1} (x_2 + x_3 ) (\phi^v_{a_1}(x_2) \phi^a_{a_1}(x_3)+ \phi^a_{a_1}(x_2)
\phi^v_{a_1}(x_3))\right] E_{nfs}(t_d)
\right\} \label{eq:b-nfs-t}\;,
\eeq
\beq
M^{N;P_1}_{nfs}&=& -\frac{32}{\sqrt{6}}\pi C_F m_{B}^2 \int_{0}^{1}d x_{1}d x_{2}\,d
x_{3}\,\int_{0}^{\infty} b_1d b_1 b_2d b_2\,
\phi_{B}(x_1,b_1)\, r_{a_1}
\non & & \times x_3 \phi^T_{a_1}(x_2) (\phi^a_{a_1}(x_3) - \phi^v_{a_1}(x_3))
\left[ h_{nfs}^c(x_1,x_2,x_3,b_1,b_2)
 E_{nfs}(t_c) \right. \non && \left.
+h_{nfs}^d(x_1,x_2,x_3,b_1,b_2) E_{nfs}(t_d)
\right]\label{eq:b-nfs-p1-n}\;,
\eeq
\beq
M_{nfs}^{T;P_1} &=& 2\, M^{N;P_1}_{nfs}\;;\label{eq:b-nf-p1-t}
\eeq
\beq
M^{N;P_2}_{nfs}&=& -\frac{32}{\sqrt{6}}\pi C_F m_{B}^2 \int_{0}^{1}d x_{1}d x_{2}\,d
x_{3}\,\int_{0}^{\infty} b_1d b_1 b_2d b_2\,
\phi_{B}(x_1,b_1)\, r_{a_1}
\non & & \times
\left\{ \left[ x_2(\phi^a_{a_1}(x_2) - \phi^v_{a_1}(x_2))
\phi^T_{a_1}(x_3)\right] h_{nfs}^d(x_1,x_2,x_3,b_1,b_2)
 \right.\non & & \left. \times E_{nfs}(t_d)
+h_{nfs}^c(x_1,x_2,x_3,b_1,b_2)
\left[x_2 (\phi^a_{a_1}(x_2) - \phi^v_{a_1}(x_2))\phi^T_{a_1}(x_3)
 \right. \right. \non && \left. \left. + 2 r_{a_1} (1- x_2 + x_3 ) (\phi^v_{a_1}(x_2) \phi^v_{a_1}(x_3)- \phi^a_{a_1}(x_2)
\phi^a_{a_1}(x_3))\right] E_{nfs}(t_c)
\right\} \label{eq:b-nfs-p2-n}\;,\\
M^{T;P_2}_{nfs}&=& -\frac{64}{\sqrt{6}}\pi C_F m_{B}^2 \int_{0}^{1}d x_{1}d x_{2}\,d
x_{3}\,\int_{0}^{\infty} b_1d b_1 b_2d b_2\,
\phi_{B}(x_1,b_1)\, r_{a_1}
\non & & \times
\left\{ \left[ x_2(\phi^a_{a_1}(x_2) - \phi^v_{a_1}(x_2))
\phi^T_{a_1}(x_3)\right] h_{nfs}^d(x_1,x_2,x_3,b_1,b_2)
 \right.\non & & \left. \times E_{nfs}(t_d)
+h_{nfs}^c(x_1,x_2,x_3,b_1,b_2)
\left[x_2 (\phi^a_{a_1}(x_2) - \phi^v_{a_1}(x_2))\phi^T_{a_1}(x_3)
 \right. \right. \non && \left. \left. + 2 r_{a_1} (1- x_2 + x_3 ) (\phi^v_{a_1}(x_2) \phi^a_{a_1}(x_3)- \phi^a_{a_1}(x_2)
\phi^v_{a_1}(x_3))\right] E_{nfs}(t_c)
\right\} \label{eq:b-nfs-p2-t}\;,
\eeq
 \beq
M_{nfa}^{N}&=& - \frac{64}{\sqrt{6}}\pi C_F m_{B}^2
\int_{0}^{1}d x_{1}d x_{2}\,d x_{3}\,\int_{0}^{\infty} b_1d b_1 b_2d
b_2\, r_{a_1}\, r_{a_1}
\non & &
 \times \phi_{B}(x_1,b_1) \left\{
h_{nfa}^f(x_1,x_2,x_3,b_1,b_2)E_{nfa}(t_f) \right. \non && \left.
\times\left[\phi_{a_1}^a(x_2) \phi^a_{a_1}(x_3)+\phi_{a_1}^v(x_2) \phi^v_{a_1}(x_3) \right]
 \right\} \label{eq:b-nfa-n}\;,\\
 M_{nfa}^{T}&=& - \frac{128}{\sqrt{6}}\pi C_F m_{B}^2
\int_{0}^{1}d x_{1}d x_{2}\,d x_{3}\,\int_{0}^{\infty} b_1d b_1 b_2d
b_2\, r_{a_1}\, r_{a_1}
\non & &
 \times \phi_{B}(x_1,b_1) \left\{
h_{nfa}^f(x_1,x_2,x_3,b_1,b_2)E_{nfa}(t_f) \right. \non && \left.
\times\left[\phi_{a_1}^v(x_2) \phi^a_{a_1}(x_3)+\phi_{a_1}^a(x_2) \phi^v_{a_1}(x_3) \right]
 \right\} \label{eq:b-nfa-t}\;,
 \eeq
 \beq
M_{nfa}^{N;P1}&=& \frac{32}{\sqrt{6}}\pi C_F m_{B}^2 \int_{0}^{1}d
x_{1}d x_{2}\,d x_{3}\,\int_{0}^{\infty} b_1d b_1 b_2d b_2\,
 \non
 & &
\times \phi_{B}(x_1,b_1)\left\{ \left[ r_{a_1}
x_2 (\phi_{a_1}^a(x_2)+\phi_{a_1}^v(x_2)) \phi^T_{a_1}(x_3) - r_{a_1} (1 -x_3)
\right.\right.
 \non
 & & \left.\left.
\times \phi^T_{a_1}(x_2)(\phi^a_{a_1}(x_3)- \phi^v_{a_1}(x_3))\right]
 E_{nfa}(t_e)h_{nfa}^e(x_1,x_2,x_3,b_1,b_2)\right. \non & & \left.
 + \left[ r_{a_1}
(2-x_2)(\phi_{a_1}^a(x_2)+\phi_{a_1}^v(x_2))
 \phi^T_{a_1}(x_3)- r_{a_1} (1+x_3)\right.\right.\non && \left.\left.\times
\phi^T_{a_1}(x_2)(\phi^a_{a_1}(x_3)-
 \phi^v_{a_1}(x_3))\right]E_{nfa}(t_f)h_{nfa}^f(x_1,x_2,x_3,b_1,b_2)
\right\} \label{eq:b-nfa-p1-n}\;,
 \eeq
 \beq
M_{nfa}^{T;P_1} &=& 2\, M_{nfa}^{N;P_1}\;; \label{eq:b-nfa-p1-t}
 \eeq
 \beq
M_{nfa}^{N;P_2} &=& M_{nfa}^{N}  \label{eq:b-nfa-p2-n}\;,\\
M_{nfa}^{T;P_2} &=& -M_{nfa}^{T} \label{eq:b-nfa-p2-t}\;;
 \eeq
 \beq
F_{fa}^{N}&=&  8 \pi C_F m_{B}^2\int_0^1 d x_{2} dx_{3}\,
\int_{0}^{\infty} b_2 db_2 b_3 db_3\,r_{a_1}\, r_{a_1}\,  \left\{ \left[
\phi^a_{a_1}(x_2) \left((x_2 +
1)\phi^a_{a_1}(x_3) \right. \right. \right. \non && \left. \left. \left. +(x_2-1)\phi^v_{a_1}(x_3)\right)+ \phi^v_{a_1}(x_2) \left((x_2 +
1)\phi^v_{a_1}(x_3)+(x_2-1)\phi^a_{a_1}(x_3)\right)\right]
E_{fa}(t_g)  \right. \non && \left.\times
 h_{fa}(x_2,1-x_3,b_2,b_3) + \left[ (x_3-2)(\phi^a_{a_1} (x_2)\phi^a_{a_1} (x_3) + \phi^v_{a_1} (x_2)\phi^v_{a_1} (x_3))
\right.\right.\non && \left.\left.- x_3 (\phi^a_{a_1} (x_2)\phi^v_{a_1} (x_3)+\phi^v_{a_1} (x_2)\phi^a_{a_1} (x_3))
 \right] E_{fa}(t_h)h_{fa}(1-x_3,x_2,b_3,b_2)
\right\}\label{eq:b-fa-n}\;,\\
F_{fa}^{T}&=&  16 \pi C_F m_{B}^2\int_0^1 d x_{2} dx_{3}\,
\int_{0}^{\infty} b_2 db_2 b_3 db_3\,r_{a_1}\, r_{a_1}\,  \left\{ \left[
\phi^v_{a_1}(x_2) \left((x_2 +
1)\phi^a_{a_1}(x_3) \right. \right. \right. \non && \left. \left. \left. +(x_2-1)\phi^v_{a_1}(x_3)\right)+ \phi^a_{a_1}(x_2) \left((x_2 +
1)\phi^v_{a_1}(x_3)+(x_2-1)\phi^a_{a_1}(x_3)\right)\right]
E_{fa}(t_g)  \right. \non && \left.\times
 h_{fa}(x_2,1-x_3,b_2,b_3) + \left[ (x_3-2)(\phi^a_{a_1} (x_2)\phi^v_{a_1} (x_3) + \phi^v_{a_1} (x_2)\phi^a_{a_1} (x_3))
\right.\right.\non && \left.\left.- x_3 (\phi^a_{a_1} (x_2)\phi^a_{a_1} (x_3)+\phi^v_{a_1} (x_2)\phi^v_{a_1} (x_3))
 \right] E_{fa}(t_h)h_{fa}(1-x_3,x_2,b_3,b_2)
\right\}\label{eq:b-fa-t}\;,
 \eeq
  \beq
 F_{fa}^{N;P_1}&=& F_{fa}^{N}\;, \label{eq:b-fa-p1-n} \\
 F_{fa}^{T;P_1}&=& F_{fa}^{T}\;, \label{eq:b-fa-p1-t}
  \eeq
  \beq
F_{fa}^{N;P_2} &=&  16 \pi C_F m_{B}^2\int_0^1 d x_{2} dx_{3}\,
\int_{0}^{\infty} b_2 db_2 b_3 db_3\,r_{a_1}\, \non && \times  \left\{ \left[
\phi^T_{a_1}(x_2) \left(\phi^a_{a_1}(x_3)- \phi^v_{a_1}(x_3)\right)\right]
E_{fa}(t_g) h_{fa}(x_2,1-x_3,b_2,b_3)
 \right. \non && \left.
 + \left[ \left(\phi^a_{a_1} (x_2)  + \phi^v_{a_1} (x_2) )
 \phi^T_{a_1} (x_3)\right)
 \right] E_{fa}(t_h)h_{fa}(1-x_3,x_2,b_3,b_2)
\right\}\label{eq:b-fa-p2-n}\;,
  \eeq
  \beq
F_{fa}^{T;P_2} &=& 2\, F_{fa}^{N;P_2}\;.\label{eq:b-fa-p2-t}
  \eeq

Thus, for these considered six tree-dominated decay channels, by combining all the possible
contributions from different Feynman diagrams, we can display the physical decay amplitudes
with three polarizations $h=L, N, T$ as follows,
\beq
{\cal M}_h(B^0 \to a_1^+ a_1^-) &=& \lambda_u \biggl[ a_1 F_{fs}^h + C_1 M_{nfs}^h + C_2 M_{nfa}^h
+ a_2 f_{B} F_{fa}^h \biggr] - \lambda_t \biggl[ (a_4+ a_{10}) F_{fs}^h  \non &&
+ (C_3 + C_9) M_{nfs}^h + (C_5+ C_7) M_{nfs}^{h;P_1} + (C_3 + 2 C_4 -\frac{1}{2}
(C_9 - C_{10})) \non  & & \times M_{nfa}^{h} + (C_5
- \frac{1}{2} C_7) M_{nfa}^{h;P_1} + (2 C_6 + \frac{1}{2} C_8) M_{nfa}^{h;P_2} + (2 a_3 + a_4
\non &&  +2 a_5 + \frac{1}{2} (a_7 + a_9 - a_{10}) ) f_B F_{fa}^{h} + (a_6
- \frac{1}{2} a_8) f_B F_{fa}^{h;P_2}   \biggr]\;, \label{eq:tda-b02a1pa1m}
\eeq
\beq
\sqrt{2} {\cal M}_{h}(B^+ \to a_1^+ a_1^0) &=& \lambda_u \biggl[(a_1 + a_2) F_{fs}^{h}
+ (C_1 + C_2) M_{nfs}^{h} \biggr] - \lambda_t \biggl[ (2 (C_9 + C_{10}) \non && -
\frac{1}{2} (3 C_7 +C_8)) F_{fs}^{h} + \frac{3}{2} (C_9 + C_{10}) M_{nfs}^{h} +
\frac{3}{2} C_7 M_{nfs}^{h;P_1} \non && + \frac{3}{2} C_8 M_{nfs}^{h;P_2} \biggr]\;, \label{eq:tda-bp2a1pa10}
\eeq
\beq
\sqrt{2} {\cal M}_{h}(B^0 \to a_1^0 a_1^0) &=& \lambda_u \biggl[ a_2
(f_B F_{fa}^{h}-F_{fs}^{h}) + C_2 (M_{nfa}^{h} - M_{nfs}^{h})\biggr] -\lambda_t \biggl[
(a_4 - \frac{1}{2} (3 a_7 \non && + 3 a_9 + a_{10}) ) F_{fs}^{h} + (C_3 - \frac{1}{2}
(C_9 + 3 C_{10})) M_{nfs}^{h} - (C_5 - \frac{1}{2} C_7) \non && \times
 M_{nfs}^{h;P_1} - \frac{3}{2}
C_8 M_{nfs}^{h;P_2} + (C_3+ 2 C_4 - \frac{1}{2} (C_9 - C_{10})) M_{nfa}^{h} \non &&  +
(C_5 - \frac{1}{2} C_7) M_{nfa}^{h;P_1} + (2 C_6 + \frac{1}{2} C_8) M_{nfa}^{h;P_2}
+ (2 a_3 + a_4 + 2 a_5 \non && + \frac{1}{2} (a_7 - a_9 + a_{10})) f_B F_{fa}^{h} + (a_6
- \frac{1}{2} a_8) f_B F_{fa}^{h;P_2}\biggr]\;. \label{eq:tda-b02a10a10}
\eeq
In the above
Eqs.~(\ref{eq:tda-b02a1pa1m})-(\ref{eq:tda-b02a10a10}),
$\lambda_u$ and $\lambda_t$ stand for the products of CKM matrix elements $V_{ub}^* V_{ud}$ and
$V_{tb}^* V_{td}$, respectively.
The standard combinations $a_i$(r.h.s.) of Wilson coefficients are defined as follows,
  \beq
a_1&=& C_2 + \frac{C_1}{3}\;, \qquad  a_2 = C_1 + \frac{C_2}{3}\;,
\qquad  a_i = C_i + \frac{C_{i \pm 1}}{3}(i=3 - 10) \;.
  \eeq
where the upper(lower) sign applies, when $i$ is odd(even).
While for $B \to b_1 b_1$ decay channels, one can easily obtain the analytic formulas for
various decay amplitudes just by replacing $a_1$ with $b_1$
in Eqs.~(\ref{eq:b-fs-l})-(\ref{eq:tda-b02a10a10}) correspondingly.

\section{Numerical Results and Discussions}\label{sec:3}

In this section, we will calculate numerically the CP-averaged
BRs, polarization fractions, direct CP-violating asymmetries, and relative phases
for those considered $B \to a_1 a_1$ and $b_1 b_1$ decay modes.
In numerical calculations, central values of the input parameters will be
used implicitly unless otherwise stated.

The QCD scale~({\rm GeV}), masses~({\rm GeV}),
 and $B$ meson lifetime({\rm ps}) are~\cite{Keum01:kpi,Amsler08:pdg}
\beq
 \Lambda_{\overline{\mathrm{MS}}}^{(f=4)} &=& 0.250\;, \qquad m_W = 80.41\;,  \qquad m_{a_1} = 1.23\;,
 \qquad m_{b_1} = 1.21\;; \non
   m_B &=& 5.279\;,\hspace{0.62cm} \quad m_b = 4.8 \;, \hspace{0.85cm} \quad   \tau_{B^+} = 1.638\;,
 \hspace{-0.10cm}  \qquad \tau_{B^0}= 1.53\;. \label{eq:mass}
\eeq

For the CKM matrix elements, we adopt the Wolfenstein
parametrization and the updated parameters $A=0.814$,
 $\lambda=0.2257$, $\bar{\rho}=0.135$, and $\bar{\eta}=0.349$~\cite{Amsler08:pdg}.

Utilizing the above chosen distribution amplitudes and the central values of the relevant
input parameters,
the resultant $B \to a_1$ and $B \to b_1$ form factors
at maximal recoil,
\beq
V_0^{B\to a_1} &=& 0.34^{+0.10}_{-0.09}  \;, \qquad
A^{B \to a_1}  = 0.27^{+0.06}_{-0.05} \;, \qquad
V_1^{B\to a_1} = 0.41^{+0.10}_{-0.08} \;;   \label{eq:form-pqcd-a1}\\
V_0^{B\to b_1} &=& 0.45^{+0.08}_{-0.09}  \;, \qquad
A^{B \to b_1}  = 0.20^{+0.05}_{-0.04} \;, \qquad
V_1^{B\to b_1} = 0.30^{+0.07}_{-0.06} \;. \label{eq:form-pqcd-b1}
\eeq
associated with the longitudinal, parallel, and perpendicular
components of the $B \to a_1 a_1$ and $B \to b_1 b_1$ decays, respectively, are
in good consistency with those as given in Ref.~\cite{Li09:afm}.

As a comparison, we quote the form factors used in the $B \to a_1 a_1$ and $b_1 b_1$ decays
in the QCD factorization~\cite{Cheng08:b2aa},
  \beq
V_0^{B\to a_1} &=& 0.30 \pm 0.05  \;, \qquad
A^{B \to a_1}  = 0.30 \pm 0.05 \;, \qquad
V_1^{B\to a_1} = 0.60 \pm 0.11 \;;   \label{eq:form-lcsr-a1}\\
V_0^{B\to b_1} &=& 0.39 \pm 0.07 \;, \qquad
A^{B \to b_1}  = 0.16 \pm 0.03 \;, \qquad
V_1^{B\to b_1} = 0.32 \pm 0.06 \;.  \label{eq:form-lcsr-b1}
\eeq
One can find that the form factors in the light-cone sum rule(LCSR), Eq.~(\ref{eq:form-lcsr-a1},\ref{eq:form-lcsr-b1}),
are basically consistent with those in the pQCD approach,
Eq.~(\ref{eq:form-pqcd-a1},\ref{eq:form-pqcd-b1}), within
errors. But, it should be noticed that, for $B \to a_1$ transition, $V_{0;{\rm LCSR}} < V_{0; {\rm pQCD}}$
in the longitudinal polarization while $A_{\rm LCSR} > A_{\rm pQCD}$ and $V_{1;{\rm LCSR}} > V_{1;{\rm pQCD}}$
in both tranverse polarizations, which may reduce the polarization fraction $f_L$, for example, for
$B^0 \to a_1^+ a_1^-$ mode evidently.

\begin{table}[t]
\caption{The CP-averaged predictions of physical observables for $B^0 \to a_1^+ a_1^-,  b_1^+ b_1^-$
decays obtained in the pQCD approach(This work).
For comparison, we also cite the available experimental measurements~\cite{Aubert09:a1pa1m}
and the theoretical estimates in the framework of
QCD factorization~\cite{Cheng08:b2aa}.} \label{tab:a1pa1m}
 \begin{center}\vspace{-0.3cm}{\tiny
\begin{tabular}[t]{c|c||c|c|c||c|c|c}
\hline  \hline
   \multicolumn{2}{c||}{Decay Channels}   &  \multicolumn{3}{c||}{$B^0 \to a_1^+ a_1^-$}& \multicolumn{3}{c}{$B^0 \to b_1^+ b_1^-$} \\
   \hline
 Parameter  & Definition & This work &   QCDF & Experiment & This work &   QCDF & Experiment\\
\hline \hline
  BR($10^{-6}$)        & $\Gamma/ \Gamma_{\rm total}$
  &$54.7^{+19.4+29.4+5.7}_{-16.9-23.2-4.4}$&
  $37.4^{+16.1+9.7}_{-13.7-1.4}$& $47.3^{+10.5+6.3}_{-10.5-6.3}$
  &$21.4^{+5.7+18.1+2.1}_{-5.3-11.3-1.4}$&
  $1.0^{+1.6+15.7}_{-0.7-0.3}$& $-$
 \\
 \hline \hline
 $f_L$      & $|{\cal A}_L|^2$
 &$0.76^{+0.01+0.03+0.00}_{-0.00-0.04-0.00}$&
  $0.64^{+0.07}_{-0.17}$& $0.31^{+0.22+0.10}_{-0.22-0.10}$
  &$0.88^{+0.02+0.04+0.01}_{-0.01-0.05-0.00}$&
  $0.96^{+0.03}_{-0.65}$& $-$
 \\
 $f_{||}$   & $|{\cal A}_{||}|^2$
 &$0.14^{+0.00+0.02+0.00}_{-0.00-0.02-0.00}$&  $-$& $-$
 &$0.07^{+0.01+0.03+0.00}_{-0.01-0.02-0.00}$&  $-$& $-$
  \\
 $f_{\perp}$& $|{\cal A}_\perp|^2$
 &$0.10^{+0.00+0.02+0.00}_{-0.00-0.01-0.00}$&  $-$& $-$
 &$0.05^{+0.01+0.02+0.00}_{-0.01-0.01-0.00}$&  $-$& $-$
 \\
 \hline \hline
 $\phi_{||}$(rad)& $\pi + \arg\frac{{\cal A}_{||}}{{\cal A}_L}$
 &$3.16^{+0.01+0.06+0.01}_{-0.01-0.04-0.01}$&  $-$& $-$
 &$2.51^{+0.07+0.07+0.01}_{-0.06-0.06-0.01}$&  $-$& $-$
 \\
 $\phi_{\perp}$(rad)& $\pi + \arg\frac{{\cal A}_{\perp}}{{\cal A}_L}$
 &$3.17^{+0.01+0.06+0.03}_{-0.00-0.04-0.01}$&  $-$& $-$
 &$2.47^{+0.06+0.08+0.01}_{-0.05-0.08-0.01}$&  $-$& $-$
 \\
 $\Delta\phi_{||}$(rad)& $\frac{\bar{\phi}_{||}-\phi_{||}}{2}$
 &$-0.04^{+0.00+0.01+0.00}_{-0.01-0.01-0.01}$&  $-$& $-$
 &$0.43^{+0.02+0.07+0.02}_{-0.04-0.09-0.04}$&  $-$& $-$
 \\
 $\Delta\phi_{\perp}$(rad)& $\frac{\bar{\phi}_\perp-\phi_\perp-\pi}{2}$
 &$-1.62^{+0.01+0.01+0.00}_{-0.01-0.01-0.01}$&  $-$& $-$
 &$-1.15^{+0.04+0.09+0.02}_{-0.02-0.09-0.03}$&  $-$& $-$
  \\
  \hline \hline
 $\acp^{\rm dir}$& $\frac{\overline{\Gamma}-\Gamma}{\overline{\Gamma}+\Gamma}$
 &$-0.04^{+0.01+0.01+0.00}_{-0.01-0.01-0.00}$&  $-$& $-$
 &$-0.00^{+0.01+0.00+0.00}_{-0.04-0.01-0.00}$&  $-$& $-$
 \\
 $\acp^{{\rm dir},L}$& $\frac{\bar{f}_L-f_L}{\bar{f}_L+f_L}$
 & $0.11^{+0.02+0.01+0.00}_{-0.02-0.01-0.01}$&  $-$& $-$
 & $-0.05^{+0.04+0.02+0.00}_{-0.04-0.03-0.00}$&  $-$& $-$
 \\
 $\acp^{{\rm dir},||}$& $\frac{\bar{f}_{||}-f_{||}}{\bar{f}_{||}+f_{||}}$
 &$-0.52^{+0.08+0.10+0.04}_{-0.07-0.07-0.02}$&  $-$& $-$
 &$0.34^{+0.05+0.02+0.01}_{-0.06-0.04-0.03}$&  $-$& $-$
 \\
 $\acp^{{\rm dir},\perp}$& $\frac{\bar{f}_\perp-f_\perp}{\bar{f}_\perp+f_\perp}$
 &$-0.54^{+0.08+0.11+0.04}_{-0.06-0.06-0.01}$& $-$& $-$
 &$0.38^{+0.06+0.03+0.02}_{-0.06-0.04-0.02}$& $-$& $-$
 \\ \hline \hline
\end{tabular}}
\end{center}
\end{table}

\subsection{CP-averaged Branching Ratios }\label{sec:3-a}

For
$B \to a_1 a_1$ and $b_1 b_1$ decays, the decay rate can
be written explicitly as,
\beq
\Gamma =\frac{G_{F}^{2}|\bf{P_c}|}{16 \pi m^{2}_{B} }
\sum_{\sigma=L,T}{\cal M}^{(\sigma)\dagger }{\cal M^{(\sigma)}}\;
\label{dr1}
\eeq
where $|\bf{P_c}|\equiv |\bf{P_{2z}}|=|\bf{P_{3z}}|$ is the momentum of either of the
outgoing axial-vector mesons and ${\cal M^{(\sigma)}}$ can be found in Eqs.~(\ref{eq:tda-b02a1pa1m})-(\ref{eq:tda-b02a10a10}).

The theoretical predictions on CP-averaged BRs for $B \to a_1 a_1$ and $b_1 b_1$ decays
evaluated in the pQCD approach,
together with the results in QCDF approach and the available experimental
data,
have been grouped in Tables~\ref{tab:a1pa1m}-\ref{tab:a10a10}.
The first error is from the B meson
wave function shape parameter $\omega_B=0.40 \pm 0.04$ GeV and decay constant
$f_B =0.19 \pm 0.02$ GeV.
The second and dominant theoretical error in these entries arises
 from the combination of the uncertainties of Gegenbauer
moments $a_{2,a_1}^{\parallel}(a_{1,b_1}^{||})$, $a_{1,a_1}^{\perp}(a_{2,b_1}^{\perp})$,
and decay constant $f_{a_1}(f_{b_1})$ in
the distribution amplitudes of axial-vector meson $a_1\, (b_1)$.
The third error is also the combined uncertainty in the CKM matrix
elements: $\bar{\rho}=0.135^{+0.031}_{-0.016} $ and
$\bar{\eta}=0.349^{+0.015}_{-0.017}$ \cite{Amsler08:pdg}. The numerical results
implicated that the input parameters(e.g. the Gegenbauer coefficients in the
nonperturbative wave functions of involved mesons) adopted in the pQCD
approach should be further improved through the better constraints from the experiments to
enhance the theoretical precision.

By comparison, one can easily
find that the results on the CP-averaged BRs of $B \to a_1 a_1$ decays in the pQCD
approach agree well with those obtained in the QCDF
within large theoretical errors,
 \beq
 \left.\begin{array}{lll}
 {\rm BR}(B^0 \to a_1^+ a_1^-) &=& 54.7^{+35.7}_{-29.0} \times 10^{-6}\;,\\
 {\rm BR}(B^+ \to a_1^+ a_1^0) &=& 21.4^{+12.6}_{-10.5} \times 10^{-6}\;, \\
 {\rm BR}(B^0 \to a_1^0 a_1^0) &=& \hspace{0.2cm}2.2^{+1.7}_{-1.1}\hspace{0.15cm} \times 10^{-6}\;; \\ \end{array} \right\}
\hspace{0.35cm}{\rm  (In \hspace{0.35cm}  pQCD) }  \label{eq:BR-a1a1-pQCD}
 \eeq
 and
 \beq
 \left.\begin{array}{lll}
 {\rm BR}(B^0 \to a_1^+ a_1^-) &=& 37.4^{+18.8}_{-13.8} \times 10^{-6}\;,\\
 {\rm BR}(B^+ \to a_1^+ a_1^0) &=& 22.4^{+12.6}_{-8.3} \times 10^{-6}\;,\\
 {\rm BR}(B^0 \to a_1^0 a_1^0) &=& \hspace{0.2cm}0.5^{+9.3}_{-0.2}\hspace{0.15cm} \times 10^{-6}\;.\\ \end{array} \right\}
\hspace{0.35cm}{\rm  (In \hspace{0.35cm}  QCDF) } \label{eq:BR-a1a1-QCDF}
 \eeq
in which various errors have been added in quadrature.

\begin{table}[t]
\caption{Same as Table~\ref{tab:a1pa1m} but
for $B^+ \to a_1^+ a_1^0, b_1^+ b_1^0$ decays.} \label{tab:a1pa10}
 \begin{center}\vspace{-0.3cm}{\tiny
\begin{tabular}[t]{c|c||c|c|c||c|c|c}
\hline  \hline
   \multicolumn{2}{c||}{Decay Channels}   &  \multicolumn{3}{c||}{$B^+ \to a_1^+ a_1^0$}& \multicolumn{3}{c}{$B^+ \to b_1^+ b_1^0$} \\
   \hline
 Parameter  & Definition & This work &   QCDF & Experiment & This work &   QCDF & Experiment\\
\hline \hline
  BR($10^{-6}$)        & $\Gamma/ \Gamma_{\rm total}$
  &$21.4^{+7.4+10.0+2.2}_{-6.5-8.2-1.0}$&
  $22.4^{+10.7+6.6}_{-8.2-1.5}$& $-$
  &$7.6^{+2.3+6.8+0.8}_{-1.9-4.1-0.6}$&
  $1.4^{+2.5+2.8}_{-1.0-0.0}$& $-$
 \\
 \hline \hline
 $f_L$      & $|{\cal A}_L|^2$
 &$0.91^{+0.00+0.02+0.00}_{-0.00-0.03-0.00}$&
  $0.74^{+0.24}_{-0.32}$& $-$
  &$0.84^{+0.01+0.05+0.00}_{-0.01-0.08-0.00}$&
  $0.95^{+0.00}_{-0.82}$& $-$
 \\
 $f_{||}$   & $|{\cal A}_{||}|^2$
 &$0.05^{+0.00+0.02+0.00}_{-0.00-0.01-0.00}$&  $-$& $-$
 &$0.10^{+0.00+0.04+0.00}_{-0.01-0.03-0.00}$&  $-$& $-$
  \\
 $f_{\perp}$& $|{\cal A}_\perp|^2$
 &$0.03^{+0.01+0.02+0.00}_{-0.00-0.00-0.00}$&  $-$& $-$
 &$0.07^{+0.00+0.03+0.00}_{-0.01-0.02-0.00}$&  $-$& $-$
 \\
 \hline \hline
 $\phi_{||}$(rad)& $\pi + \arg\frac{{\cal A}_{||}}{{\cal A}_L}$
 &$3.28^{+0.02+0.08+0.00}_{-0.02-0.09-0.00}$&  $-$& $-$
 &$2.75^{+0.06+0.06+0.00}_{-0.05-0.05-0.00}$&  $-$& $-$
 \\
 $\phi_{\perp}$(rad)& $\pi + \arg\frac{{\cal A}_{\perp}}{{\cal A}_L}$
 &$3.28^{+0.02+0.09+0.00}_{-0.02-0.10-0.00}$&  $-$& $-$
 &$2.77^{+0.06+0.07+0.00}_{-0.05-0.05-0.00}$&  $-$& $-$
 \\
 $\Delta\phi_{||}$(rad)& $\frac{\bar{\phi}_{||}-\phi_{||}}{2}$
 &$\approx 0.0$&  $-$& $-$
 &$\approx 0.0$&  $-$& $-$
 \\
 $\Delta\phi_{\perp}$(rad)& $\frac{\bar{\phi}_\perp-\phi_\perp-\pi}{2}$
 &$\approx -1.58$&  $-$& $-$
  &$\approx -1.57$&  $-$& $-$
  \\
  \hline \hline
 $\acp^{\rm dir}$& $\frac{\overline{\Gamma}-\Gamma}{\overline{\Gamma}+\Gamma}$
 &$\approx 0.0$&  $-$& $-$
 &$\approx 0.0$&  $-$& $-$
 \\
 $\acp^{{\rm dir},L}$& $\frac{\bar{f}_L-f_L}{\bar{f}_L+f_L}$
 & $\approx 0.0$&  $-$& $-$
 &$\approx 0.0$&  $-$& $-$
 \\
 $\acp^{{\rm dir},||}$& $\frac{\bar{f}_{||}-f_{||}}{\bar{f}_{||}+f_{||}}$
 &$\approx 0.0$&  $-$& $-$
 &$\approx 0.0$&  $-$& $-$
 \\
 $\acp^{{\rm dir},\perp}$& $\frac{\bar{f}_\perp-f_\perp}{\bar{f}_\perp+f_\perp}$
 &$\approx 0.0$& $-$& $-$
 &$\approx 0.0$&  $-$& $-$
 \\ \hline \hline
\end{tabular}}
\end{center}
\end{table}

Based on those numerical results given in Tables~\ref{tab:a1pa1m}-\ref{tab:a10a10},
some remarks on the CP-averaged BRs for $B \to a_1 a_1$ decays are in order:
\begin{enumerate}
\item
As mentioned in the introduction, the experimental measurement for $B^0 \to a_1^+ a_1^-$ mode has been performed
by BaBar Collaboration and the decay rate is~\cite{Aubert09:a1pa1m,Asner10:hfag},
 \beq
{\rm BR}(B^0 \to a_1^+ a_1^-)_{\rm Exp.} &=& 47.3^{+12.2}_{-12.2} \times 10^{-6}\;.\label{eq:val-expt}
 \eeq
Combined with the numerical results evaluated with pQCD and QCDF approaches\footnote{Here,
we do not quote BR$(B^0 \to a_1^+ a_1^-)$
provided in naive factorization~\cite{Calderon07:b2aa}
for phenomenological analysis
just because the numerical result is too small to be comparable with the data.}, i.e.,
Eqs.~(\ref{eq:BR-a1a1-pQCD}) and (\ref{eq:BR-a1a1-QCDF}),
one can find that the predicted BR$(B^0 \to a_1^+ a_1^-)_{\rm Th.}$
agree well with the present data within uncertainties.

\item
Based on the discussions~\cite{Yang07:twist,Cheng07:b2ap,Cheng08:b2aa} about
the hadron dynamics
of axial-vector $a_1$ and vector $\rho$
mesons, due to the similar QCD behavior between them, the decay pattern
of $B \to a_1 a_1$ modes is therefore analogous to that of $B \to \rho \rho$ ones as expected.

\item
Because the decay constant $f_{a_1} \approx 1.2 \times f_{\rho}$
$\approx 1.5 \times f_{\rho}^T$, the pattern of ${\rm BR}(B \to
a_1 a_1) > {\rm BR}(B \to \rho \rho)$ is also as expected in the pQCD approach correspondingly.
But, it is worth mentioning that for the color-suppressed decay $B^0 \to a_1^0 a_1^0$, ${\rm BR}
(B^0 \to a_1^0 a_1^0)_{\rm pQCD} \approx 4 \times {\rm BR}
(B^0 \to a_1^0 a_1^0)_{\rm QCDF}$.

\item
Actually, in view of the larger decay constant $f_{a_1}$ and
similar QCD behavior of $a_1$ to $\rho$, as a naive estimate, the relation
of the BRs among three channels $B^0 \to \rho^0 \rho^0$, $B^0 \to a_1^0 \rho^0$,
and $B^0 \to a_1^0\, a_1^0$ may
be ${\rm BR}(B^0 \to \rho^0 \rho^0) < {\rm BR}(B^0 \to a_1^0 \rho^0)
< {\rm BR}(B^0 \to a_1^0 a_1^0)$. In other words, ${\rm BR}
(B^0 \to a_1^0\, a_1^0)_{\rm QCDF}$ should be comparable with 
${\rm BR}(B^0 \to a_1^0\, a_1^0)_{\rm pQCD}$.
It is thus a bit strange that ${\rm BR}(B^0 \to \rho^0 \rho^0)
\approx 2 \times {\rm BR}(B^0 \to a_1^0 a_1^0)$ and ${\rm BR}(B^0 \to a_1^0 \rho^0)
\approx 2.5 \times {\rm BR}(B^0 \to a_1^0 a_1^0)$ in Ref.~\cite{Cheng08:b2aa} within the framework of QCDF. It is worth stressing that among the above comparisons, the central values of the
theoretical predictions are adopted for clarification.

\item
Similar to $B \to \pi \pi$ and $B \to \rho \rho$ decays, one can find that
BR($B^0 \to a_1^+ a_1^-$)$ >$ BR($B^+ \to a_1^+ a_1^0$) $>$ BR($B^0 \to a_1^0 a_1^0$) in the
pQCD approach.
Moreover, like $B^0 \to a_1^\pm \pi^\mp$ decay~\cite{Wang08:b2ap}, in principle, $B^0 \to a_1^+ a_1^-$ can be utilized to provide an independent
measurement of the CKM angle $\alpha$~\cite{Lombardo09:b2aa}. Certainly, the currently available statistics is
too low to perform such an analysis on the angle of $\alpha$ 
from the current generation of B factories.
But, potentially, a new generation of Super flavor factories(SuperB) are
expected to achieve the result with a high luminosity
$\gsim 10^{36} {\rm cm}^{-2} {\rm s}^{-1}$~\cite{Soni07,SuperB}.

\end{enumerate}

\begin{table}[b]
\caption{Same as Table~\ref{tab:a1pa1m} but
for $B^0 \to a_1^0\, a_1^0\,, b_1^0\, b_1^0$ decays.} \label{tab:a10a10}
 \begin{center}\vspace{-0.3cm}{\tiny
\begin{tabular}[t]{c|c||c|c|c||c|c|c}
\hline  \hline
   \multicolumn{2}{c||}{Decay Channels}   &  \multicolumn{3}{c||}{$B^0 \to a_1^0\, a_1^0$}& \multicolumn{3}{c}{$B^0 \to b_1^0\, b_1^0$} \\
   \hline
 Parameter  & Definition & This work &   QCDF & Experiment & This work &   QCDF & Experiment\\
\hline \hline
  BR($10^{-6}$)        & $\Gamma/ \Gamma_{\rm total}$
  &$2.2^{+0.6+1.6+0.1}_{-0.5-1.0-0.1}$&
  $0.5^{+0.8+9.3}_{-0.2-0.0}$& $-$
  &$29.0^{+7.7+27.9+3.1}_{-6.9-16.6-2.2}$&
  $3.2^{+5.6+11.0}_{-2.3-0.8}$& $-$
 \\
 \hline \hline
 $f_L$      & $|{\cal A}_L|^2$
 &$0.12^{+0.01+0.02+0.00}_{-0.01-0.02-0.01}$&
  $0.60^{+0.00}_{-0.70}$& $-$
  &$\approx 1.00$&
  $0.95^{+0.02}_{-0.80}$& $-$
 \\
 $f_{||}$   & $|{\cal A}_{||}|^2$
 &$0.49^{+0.00+0.00+0.00}_{-0.01-0.01-0.00}$&  $-$& $-$
 &$\approx 0.00$&  $-$& $-$
  \\
 $f_{\perp}$& $|{\cal A}_\perp|^2$
 &$0.40^{+0.00+0.01+0.00}_{-0.01-0.02-0.01}$&  $-$& $-$
 &$\approx 0.00$&  $-$& $-$
 \\
 \hline \hline
 $\phi_{||}$(rad)& $\pi + \arg\frac{{\cal A}_{||}}{{\cal A}_L}$
 &$3.58^{+0.05+0.01+0.06}_{-0.06-0.92-0.05}$&  $-$& $-$
 &$4.20^{+0.03+0.09+0.04}_{-0.02-0.19-0.00}$&  $-$& $-$
 \\
 $\phi_{\perp}$(rad)& $\pi + \arg\frac{{\cal A}_{\perp}}{{\cal A}_L}$
 &$3.64^{+0.07+0.02+0.06}_{-0.07-0.99-0.04}$&  $-$& $-$
 &$4.24^{+0.03+0.08+0.04}_{-0.01-0.15-0.00}$&  $-$& $-$
 \\
 $\Delta\phi_{||}$(rad)& $\frac{\bar{\phi}_{||}-\phi_{||}}{2}$
 &$0.77^{+0.10+0.02+0.07}_{-0.07-1.52-0.09}$&  $-$& $-$
 &$-0.32^{+0.08+0.16+0.09}_{-0.06-0.24-0.00}$&  $-$& $-$
 \\
 $\Delta\phi_{\perp}$(rad)& $\frac{\bar{\phi}_\perp-\phi_\perp-\pi}{2}$
 &$-0.83^{+0.08+0.02+0.06}_{-0.06-1.50-0.08}$&  $-$& $-$
 &$-1.85^{+0.06+0.13+0.07}_{-0.05-0.19-0.00}$&  $-$& $-$
  \\
  \hline \hline
 $\acp^{\rm dir}$& $\frac{\overline{\Gamma}-\Gamma}{\overline{\Gamma}+\Gamma}$
 &$-0.78^{+0.07+0.00+0.03}_{-0.04-0.00-0.01}$&  $-$& $-$
 &$-0.03^{+0.02+0.00+0.02}_{-0.01-0.00-0.01}$&  $-$& $-$
 \\
 $\acp^{{\rm dir},L}$& $\frac{\bar{f}_L-f_L}{\bar{f}_L+f_L}$
 &$0.20^{+0.09+0.05+0.00}_{-0.55-0.05-0.01}$&  $-$& $-$
 &$-0.03^{+0.02+0.00+0.02}_{-0.02-0.00-0.01}$&  $-$& $-$
 \\
 $\acp^{{\rm dir},||}$& $\frac{\bar{f}_{||}-f_{||}}{\bar{f}_{||}+f_{||}}$
 &$-0.91^{+0.09+0.00+0.04}_{-0.04-0.00-0.01}$&  $-$& $-$
 &$0.19^{+0.00+0.30+0.00}_{-0.10-0.34-0.15}$&  $-$& $-$
 \\
 $\acp^{{\rm dir},\perp}$& $\frac{\bar{f}_\perp-f_\perp}{\bar{f}_\perp+f_\perp}$
 &$-0.90^{+0.10+0.00+0.04}_{-0.04-0.00-0.01}$& $-$& $-$
 &$0.23^{+0.01+0.30+0.00}_{-0.10-0.33-0.15}$& $-$& $-$
 \\ \hline \hline
\end{tabular}}
\end{center}
\end{table}

Now let's turn to analyze the phenomenologies of $B \to b_1 b_1$ decays.
One can find that the pQCD
predictions for BRs of $B \to b_1 b_1$ decays are
\beq
 \left.\begin{array}{lll}
 {\rm BR}(B^0 \to b_1^+ b_1^-) &=& 21.4^{+19.1}_{-12.6} \times 10^{-6}\;,\\
 {\rm BR}(B^+ \to b_1^+ b_1^0) &=& \hspace{0.35cm}7.6^{+7.2}_{-4.6} \times 10^{-6}\;, \\
 {\rm BR}(B^0 \to b_1^0 b_1^0) &=& 29.0^{+29.1}_{-18.1} \times 10^{-6}\;; \\ \end{array} \right\}
\hspace{0.35cm}{\rm  (In \hspace{0.35cm}  pQCD) } \label{eq:BR-b1b1-pQCD}
 \eeq
 and that in the QCDF approach,
 \beq
 \left.\begin{array}{lll}
 {\rm BR}(B^0 \to b_1^+ b_1^-) &=& 1.0^{+15.8}_{-0.8} \times 10^{-6}\;,\\
 {\rm BR}(B^+ \to b_1^+ b_1^0) &=& \hspace{0.16cm}1.4^{+3.8}_{-1.0} \times 10^{-6}\;,\\
 {\rm BR}(B^0 \to b_1^0 b_1^0) &=& 3.2^{+12.3}_{-2.4} \times 10^{-6}\;.\\ \end{array} \right\}
\hspace{0.35cm}{\rm  (In \hspace{0.35cm}  QCDF) }  \label{eq:BR-b1b1-QCDF}
 \eeq
in which various errors have been added in quadrature too. Though the central values
of the CP-averaged BRs for $B \to b_1 b_1$ decays in the pQCD approach are much larger
than those in QCD factorization, they are roughly consistent with each other within large errors and
in the order of $10^{-6} \sim 10^{-5}$. It will be of great interest to measure these
theoretical predictions to test pQCD and QCDF approaches experimentally.

\begin{table}[t]
\caption{ The decay amplitudes(in unit of $10^{-3}\; \rm{GeV}^3$) of the
$B^0 \to a_1^+ a_1^-, b_1^+ b_1^-$ channels
with three polarizations, where only the central values are quoted for clarification.}
\label{tab:DA-pm}
 \begin{center}\vspace{-0.3cm}{\tiny
\begin{tabular}[t]{c||c|c|c|c|c|c|c|c}
\hline  \hline
  Channel   &  \multicolumn{8}{c}{$B^0 \to a_1^+ a_1^-$}\\
   \hline
 Decay Amplitudes & ${\cal A}^T_{fs}$ & ${\cal A}^P_{fs}$ &  ${\cal A}^T_{nfs}$ &  ${\cal A}^P_{nfs}$
                  & ${\cal A}^T_{nfa}$ &  ${\cal A}^P_{nfa}$ & ${\cal A}^T_{fa}$  & ${\cal A}^P_{fa}$\\
\hline \hline
 $L$      &$ 3.23 +{\it i} 8.36$ &  $-0.82 + {\it i} 0.34$
          &$-0.16 -{\it i} 0.33$ &  $ 0.03 - {\it i} 0.01$
          &$ 0.47 +{\it i} 0.09$ &  $-0.11 + {\it i} 0.29$
          &$ \sim 0.00 $              &  $ 0.31 + {\it i} 0.21$
 \\
 $N$      &$ 0.76 +{\it i} 1.98$ &  $-0.19 + {\it i} 0.08$
          &$ 0.28 +{\it i} 0.33$ &  $-0.03 + {\it i} 0.03$
          &$ 0.02 +{\it i} 0.01$ &  $-0.00 + {\it i} 0.01$
          &$ \sim 0.00$               &  $-0.41 - {\it i} 0.61$
 \\
 $T$      &$ 1.46 +{\it i} 3.77$ &  $-0.36 + {\it i} 0.15$
          &$ 0.57 +{\it i} 0.71$ &  $-0.06 + {\it i} 0.05$
          &$ \sim 0.00$               &  $ \sim 0.00$
          &$-0.02 -{\it i} 0.00$ &  $ 0.81 - {\it i} 1.21$
 \\
 \hline \hline
  Channel  & \multicolumn{8}{c}{$B^0 \to b_1^+ b_1^-$} \\
   \hline
 Decay Amplitudes & ${\cal A}^T_{fs}$ & ${\cal A}^P_{fs}$ &  ${\cal A}^T_{nfs}$ &  ${\cal A}^P_{nfs}$
                  & ${\cal A}^T_{nfa}$ &  ${\cal A}^P_{nfa}$ & ${\cal A}^T_{fa}$  & ${\cal A}^P_{fa}$\\
\hline \hline
 $L$      &$-0.01 -{\it i} 0.02$ &  $ \sim 0.00$
          &$-1.10 +{\it i} 3.50$ &  $-0.36 - {\it i} 0.09$
          &$ 0.52 +{\it i} 1.89$ &  $-1.26 + {\it i} 0.42$
          &$-0.01 -{\it i} 0.00$ &  $-1.03 - {\it i} 0.73$
 \\
 $N$      &$ 0.43 +{\it i} 1.12$ &  $-0.11 + {\it i} 0.05$
          &$ 0.14 -{\it i} 0.12$ &  $ 0.01 + {\it i} 0.02$
          &$-0.02 +{\it i} 0.01$ &  $ \sim 0.00$
          &$ \sim 0.00$               &  $ 0.17 + {\it i} 0.13$
 \\
 $T$      &$ 0.81 +{\it i} 2.09$ &  $-0.21 + {\it i} 0.09$
          &$ 0.32 -{\it i} 0.21$ &  $ 0.02 + {\it i} 0.04$
          &$ \sim 0.00$               &  $ \sim 0.00$
          &$ 0.01 -{\it i} 0.00$ &  $ 0.35 + {\it i} 0.29$
 \\
 \hline \hline
\end{tabular}}
\end{center}
\end{table}

From the theoretical predictions presented in Tables~\ref{tab:a1pa1m}-\ref{tab:a10a10},
some discussions on the CP-averaged BRs for $B \to b_1 b_1$ decays
are given as follows:

\begin{enumerate}
\item
Because of charge conjugation
invariance, the decay constant of the $^1P_1$ neutral meson $b^0_1$
 must be zero. In the isospin
limit, the decay constant of the charged $b_1$ vanishes due
to the fact that the $b_1$ has even G-parity and that the
relevant weak axial-vector current is odd under G transformation.
Hence, $f_{b^\pm_1}$
is very small in reality. In the numerical calculations, we adopted
$f_{b^\pm_1} = \mp (0.0028 \pm 0.0026) f_{b_1}$~\cite{Cheng08:b2aa}.

\item
In principle, because of either very small or vanishing decay constant
of meson $b_1$, the BRs of $B \to b_1 b_1$ decays should be significantly
suppressed by comparing with the $B \to a_1 a_1$ ones as naive anticipation. However,
as shown in Eqs.~(\ref{eq:wfl-a})-(\ref{eq:inputs}), the $b_1 (^1P_1)$ meson
has the rather different hadron dynamics from its $^3P_1$ partner, the $a_1$ meson.
Correspondingly, one can
find the induced anomaly in the theoretical pQCD
predictions on the CP-averaged BRs and other physical observables shown in Tables~\ref{tab:a1pa1m}-\ref{tab:a10a10}.

\item
The $B \to b_1 b_1$ decays receive
dominantly large contributions arising from the nonfactorizable spectator 
diagrams, which result in the large CP-averaged BRs. To clarify this point more clearly,
we present the decay amplitudes numerically for every topology with three polarizations
in the Tables~\ref{tab:DA-pm}-\ref{tab:DA-00}, where only the central values are quoted.

\item
Moreover, the BRs of
the $B \to b_1 b_1$ decays exhibit an interesting pattern highly different from that of $B \to
a_1 a_1$ ones. In terms of the central values in the pQCD approach as listed in Tables~\ref{tab:a1pa1m}-\ref{tab:a10a10},
one can observe that ${\rm BR}(B^0 \to b_1^0 b_1^0)
> {\rm BR}(B^0 \to b_1^+ b_1^-) > {\rm BR}(B^+ \to b_1^+ b_1^0)$, while
${\rm BR}(B^0 \to a_1^0 a_1^0) < {\rm BR}(B^+ \to a_1^+ a_1^0) < {\rm BR}(B^0 \to a_1^+ a_1^-)$.
Meanwhile, one can also find that ${\rm BR}(B^0 \to b_1^0 b_1^0)
> {\rm BR}(B^+ \to b_1^+ b_1^0) > {\rm BR}(B^0 \to b_1^+ b_1^-)$,
while ${\rm BR}(B^0 \to a_1^0 a_1^0)
< {\rm BR}(B^+ \to a_1^+ a_1^0) < {\rm BR}(B^0 \to a_1^+ a_1^-)$ in the QCDF approach~\cite{Cheng08:b2aa}.
The confirmation of these interesting relations through the relevant experiments may shed light
on the QCD dynamics involved in these considered channels.

\item
Although the components of $B \to b_1 b_1$ decays at the quark level are same as
that of $B \to \rho \rho$ and $B \to a_1 a_1$ decays, the different phenomenologies have been
shown(See Tables~\ref{tab:a1pa1m}-\ref{tab:a10a10}) because of the different QCD behavior
between $a_1$ and $b_1$ mesons.
One can therefore
expect that $B \to b_1 b_1$ decays
will imply some new information on the CKM unitary angle, reliability of pQCD approach, and so on.

\end{enumerate}


\begin{table}[t]
\caption{ The decay amplitudes(in unit of $10^{-3}\; \rm{GeV}^3$) of the
$B^+ \to a_1^+ a_1^0, b_1^+ b_1^0$ channels
with three polarizations, where only the central values are quoted for clarification.}
\label{tab:DA-p0}
 \begin{center}\vspace{-0.3cm}{\tiny
\begin{tabular}[t]{c||c|c|c|c|c|c|c|c}
\hline  \hline
  Channel   &  \multicolumn{8}{c}{$B^+ \to a_1^+ a_1^0$}\\
   \hline
 Decay Amplitudes & ${\cal A}^T_{fs}$ & ${\cal A}^P_{fs}$ &  ${\cal A}^T_{nfs}$ &  ${\cal A}^P_{nfs}$
                  & ${\cal A}^T_{nfa}$ &  ${\cal A}^P_{nfa}$ & ${\cal A}^T_{fa}$  & ${\cal A}^P_{fa}$\\
\hline \hline
 $L$      &$ 1.98 +{\it i} 5.13$ &$-0.18 + {\it i} 0.07$
          &$ 0.18 +{\it i} 0.34$ &$-0.01 + {\it i} 0.00$
          &$-$ &  $-$  &$-$ &  $-$
 \\
 $N$      &$ 0.48 +{\it i} 1.25$ &  $-0.04 + {\it i} 0.02$
          &$-0.25 -{\it i} 0.27$ &  $ 0.01 - {\it i} 0.01$
          &$-$ &  $-$  &$-$ &  $-$
 \\
 $T$      &$ 0.92 +{\it i} 2.37$ &  $-0.08 +{\it i} 0.03$
          &$-0.49 -{\it i} 0.60$ &  $ 0.01 -{\it i} 0.01$
          &$-$ &  $-$  &$-$ &  $-$
 \\
 \hline \hline
  Channel  & \multicolumn{8}{c}{$B^+ \to b_1^+ b_1^0$} \\
   \hline
 Decay Amplitudes & ${\cal A}^T_{fs}$ & ${\cal A}^P_{fs}$ &  ${\cal A}^T_{nfs}$ &  ${\cal A}^P_{nfs}$
                  & ${\cal A}^T_{nfa}$ &  ${\cal A}^P_{nfa}$ & ${\cal A}^T_{fa}$  & ${\cal A}^P_{fa}$\\
\hline \hline
 $L$      &$-0.01 -{\it i} 0.01$ &  $ \sim 0.00$
          &$ 0.59 -{\it i} 3.30$ &  $ 0.11 - {\it i} 0.02$
          &$-$ &  $-$  &$-$ &  $-$
 \\
 $N$      &$ 0.26 +{\it i} 0.68$ &  $-0.02 + {\it i} 0.01$
          &$-0.09 +{\it i} 0.10$ &  $ \sim 0.00$
          &$-$ &  $-$  &$-$ &  $-$
 \\
 $T$      &$ 0.49 +{\it i} 1.27$ &  $-0.04 + {\it i} 0.02$
          &$-0.20 +{\it i} 0.20$ &  $-0.01 - {\it i} 0.01$
          &$-$ &  $-$  &$-$ &  $-$
 \\
 \hline \hline
\end{tabular}}
\end{center}
\end{table}

As mentioned in the above,
our pQCD prediction on the BR of $B^0 \to a_1^+ a_1^-$ decay
is consistent with the data reported by BaBar Collaboration very recently.
Though the theoretical errors are a bit large, the central value of this channel can still
be employed to roughly estimate the BRs of other decay modes.
Here we define four parameters, say,
"${\rm R}_{aa}^{0+}$", "${\rm R}_{bb}^{0+}$", "${\rm R}_{ab}^{00}$" and "${\rm R}_{ab}^{++}$",
as follows,
\beq
{\rm R}_{aa}^{0+} &\equiv& \frac{\tau_{B^+}}{\tau_{B^0}} \cdot
\frac{{\rm BR}(B^0 \to a_1^+ a_1^-)}{{\rm BR}(B^+ \to a_1^+ a_1^0)}
 \approx 2.7\;,\quad
{\rm R}_{bb}^{0+} \equiv \frac{\tau_{B^+}}{\tau_{B^0}} \cdot
\frac{{\rm BR}(B^0 \to b_1^+ b_1^-)}{{\rm BR}(B^+ \to b_1^+ b_1^0)}
 \approx 3.0\;;
\eeq
 \beq
{\rm R}_{ab}^{00} &\equiv&
\frac{{\rm BR}(B^0 \to a_1^+ a_1^-)}{{\rm BR}(B^0 \to b_1^+ b_1^-)}
 \approx 2.6\;,\qquad \qquad
{\rm R}_{ab}^{++} \equiv
\frac{{\rm BR}(B^+ \to a_1^+ a_1^0)}{{\rm BR}(B^+ \to b_1^+ b_1^0)}
 \approx 2.8\;.
\eeq
We expect the above four ratios could be tested at the ongoing LHC and
forthcoming Super-B experiments.

Finally, we should stress that both of the color-suppressed modes $B^0 \to a_1^0 a_1^0$
and $B^0 \to b_1^0 b_1^0$ themselves exhibit the dramatically different phenomenologies.
Similar to $B^0 \to \pi^0 \pi^0, \pi^0 \rho^0, \rho^0 \rho^0$ decays\footnote{As for the
color-suppressed processes in the decays of $B$ mesons,
which have been extensively
studied in plenties of literatures with various of methods and/or schemes within and
beyond the standard model.
However, to our best knowledge, they seem to be a longstanding "puzzle" in $B$ physics
because one can not resolve
it self-consistently in the current approaches/methods.},
$B^0 \to a_1^0 a_1^0$ and $b_1^0 b_1^0$ decays are closely
related to the color-suppressed tree amplitudes $C$~\footnote{Unfortunately, up to now,
the color-suppressed tree amplitude $C$ seems to be an important
but the least understood quantity in $B$ meson decays~\cite{Li11:puzzle}}
$\propto a_2 (= C_1 + {C_2}/3)$, where $C_1$ and $C_2$ are Wilson coefficients.
At leading order, the sign of $C_2$ is positive while the sign of $C_1$ is negative,
which can cancel each other mostly. For example, the numerical result of $a_2$ is about $1.1 \times 10^{-3}$
when the running hard scale is taken at $\mu = 2.5$ GeV~\cite{Lu07:bs2mm}.
Furthermore, one can easily find from the Table~\ref{tab:DA-00} that the tree contributions
${\cal A}^T_{fs}$ and ${\cal A}^T_{nfs}$ from factorizable spectator and
nonfactorizable spectator diagrams cancel each other significantly for $B^0 \to a_1^0 a_1^0$ channel.
Thus BR($B^0 \to a_1^0 a_1^0$)
is rather small relative to BR($B^0 \to a_1^+ a_1^-$) and BR($B^+ \to a_1^+ a_1^0$)~\footnote{Recently,
the authors in Ref.~\cite{Li11:puzzle} proposed a solution to the $B \to \pi \pi$ puzzle by
considering the contributions arising from "Glauber-gluon-region". However, it is worth mentioning
that this class of contributions may make very little effects to the results on the considered
$B \to a_1^0 a_1^0, b_1^0 b_1^0$ decays in the present work, which because, as argued
in the literature, the Glauber effects can only contribute significantly to the pion but much less
to the $\rho$ meson.}.
It should be stressed that the small quantity $a_2$ in $B^0 \to b_1^0 b_1^0$ need not to be
considered seriously because the decay constant $f_{b_1^0}$ is exact zero. But,
the resulting BR($B^0 \to b_1^0 b_1^0$)
is such large that reaching $29.0 \times 10^{-6}$ numerically. It will be highly interesting
to measure this rate to test the availability of pQCD approach in the channels with $^1P_1$ mesons.

\begin{table}[t]
\caption{ The decay amplitudes(in unit of $10^{-3}\; \rm{GeV}^3$) of the
$B^0 \to a_1^0\; a_1^0, b_1^0\; b_1^0$ channels
with three polarizations, where only the central values are quoted for clarification.}
\label{tab:DA-00}
 \begin{center}\vspace{-0.3cm}{\tiny
\begin{tabular}[t]{c||c|c|c|c|c|c|c|c}
\hline  \hline
  Channel   &  \multicolumn{8}{c}{$B^0 \to a_1^0\, a_1^0$}\\
   \hline
 Decay Amplitudes & ${\cal A}^T_{fs}$ & ${\cal A}^P_{fs}$ &  ${\cal A}^T_{nfs}$ &  ${\cal A}^P_{nfs}$
                  & ${\cal A}^T_{nfa}$ &  ${\cal A}^P_{nfa}$ & ${\cal A}^T_{fa}$  & ${\cal A}^P_{fa}$\\
\hline \hline
 $L$      &$ 0.30 +{\it i} 0.79$ &  $-0.40 + {\it i} 0.17$
          &$-0.28 -{\it i} 0.58$ &  $ 0.03 - {\it i} 0.01$
          &$ 0.31 +{\it i} 0.07$ &  $-0.08 + {\it i} 0.20$
          &$ \sim 0.00$ &  $ 0.22 + {\it i} 0.14$
 \\
 $N$      &$ 0.06 +{\it i} 0.15$ &  $-0.09 + {\it i} 0.04$
          &$ 0.45 +{\it i} 0.51$ &  $-0.03 + {\it i} 0.02$
          &$ 0.02 +{\it i} 0.01$ &  $ \sim 0.00$
          &$ \sim 0.00$ &  $-0.29 - {\it i} 0.43$
 \\
 $T$      &$ 0.12 +{\it i} 0.30$ &  $-0.17 + {\it i} 0.07$
          &$ 0.88 +{\it i} 1.10$ &  $-0.06 + {\it i} 0.05$
          &$ \sim 0.00$ &  $ \sim 0.00$
          &$-0.01 -{\it i} 0.00$ &  $-0.59 - {\it i} 0.86$
 \\
 \hline \hline
  Channel  & \multicolumn{8}{c}{$B^0 \to b_1^0\, b_1^0$} \\
   \hline
 Decay Amplitudes & ${\cal A}^T_{fs}$ & ${\cal A}^P_{fs}$ &  ${\cal A}^T_{nfs}$ &  ${\cal A}^P_{nfs}$
                  & ${\cal A}^T_{nfa}$ &  ${\cal A}^P_{nfa}$ & ${\cal A}^T_{fa}$  & ${\cal A}^P_{fa}$\\
\hline \hline
 $L$      &$ 0.00$ &  $ 0.00$
          &$-1.34 +{\it i} 5.77$ &  $-0.30 - {\it i} 0.12$
          &$ 0.34 +{\it i} 1.32$ &  $-0.90 + {\it i} 0.31$
          &$ \sim 0.00$ &  $-0.73 - {\it i} 0.52$
 \\
 $N$      &$ 0.04 +{\it i} 0.11$ &  $-0.05 + {\it i} 0.02$
          &$ 0.19 -{\it i} 0.18$ &  $ 0.01 + {\it i} 0.01$
          &$-0.01 +{\it i} 0.01$ &  $ \sim 0.00$
          &$ \sim 0.00$ &  $ 0.12 + {\it i} 0.10$
 \\
 $T$      &$ 0.08 +{\it i} 0.21$ &  $-0.10 + {\it i} 0.04$
          &$ 0.42 -{\it i} 0.35$ &  $ 0.02 + {\it i} 0.03$
          &$ \sim 0.00$ &  $ \sim 0.00$
          &$ 0.01 -{\it i} 0.00$ &  $ 0.25 + {\it i} 0.20$
 \\
 \hline \hline
\end{tabular}}
\end{center}
\end{table}

\subsection{Polarization Fractions}\label{sec:3-b}

We have also computed the polarization fractions for $B \to a_1 a_1$ and $b_1 b_1$ decay modes.
Based on the helicity amplitudes~(\ref{eq:amp}), we can define the
transversity amplitudes as
\beq
{\cal A}_{L}&=&-\xi
m^{2}_{B}{\cal M}_{L}, \quad {\cal A}_{\parallel}=\xi
\sqrt{2}m^{2}_{B}{\cal M}_{N}, \quad {\cal A}_{\perp}=\xi r_2 r_3
\sqrt{2(r^{2}-1)} m^{2}_{B} {\cal M }_{T}\;. \label{eq:ase}
\eeq
for the longitudinal, parallel, and perpendicular polarizations,
respectively, with the normalization factor
$\xi=\sqrt{G^2_{F}{\bf{P_c}} /(16\pi m^2_{B}\Gamma)}$ and the
ratio $r=P_{2}\cdot P_{3}/(m_{B}^2\; r_2 r_3)$. These amplitudes satisfy
the relation, \beq |{\cal A}_{L}|^2+|{\cal
A}_{\parallel}|^2+|{\cal A}_{\perp}|^2=1\;. \eeq following the
summation in Eq.~(\ref{dr1}).
The polarization fractions $f_{L},f_{||}$ and $f_{\perp}$ can thus be
read as,
\beq
f_{L(||,\perp)} &\equiv& \frac{|{\cal
A}_{L(||,\perp)}|^2}{|{\cal A}_L|^2+|{\cal A}_{||}|^2+|{\cal
A}_{\perp}|^2} = |{\cal A}_{L(||,\perp)}|^2, \label{eq:pfs}
\eeq

The numerical results of fractions with three polarizations for $B \to a_1 a_1$
and $B \to b_1 b_1$ decays in the pQCD approach have been presented in
Tables~\ref{tab:a1pa1m}-\ref{tab:a10a10}. Based on these values, we give some
phenomenological analysis:
\begin{enumerate}
\item
As discussed in Sec.~\ref{sec:3-a}, the prediction on BR$(B^0 \to a_1^+ a_1^-)$
with pQCD and QCDF approach, respectively, is in good agreement with the measurement given by
BaBar collaboration. However, the fraction of longitudinal polarization
for $B^0 \to a_1^+ a_1^-$ is not the case. Theoretically, this considered channel is
dominated by the longitudinal contributions,
   \beq
   f_L(B^0 \to a_1^+\, a_1^-)_{\rm pQCD}&=& 0.76^{+0.03}_{-0.04}\;,\\
   f_L(B^0 \to a_1^+\, a_1^-)_{\rm QCDF}&=& 0.64^{+0.07}_{-0.17}\;;
   \eeq
and $f_L$ predicted in pQCD and QCDF
are very close to each other;
on the other hand, experimentally,
 \beq
 f_L(B^0 \to a_1^+\, a_1^-)_{\rm Exp.} &=& 0.31^{+0.24}_{-0.24}\;. \label{eq:longf-expt.}
 \eeq
it seems to be governed by the transverse ones. But, it should be mentioned that the measurement performed by
BaBar collaboration still have very large errors and should be
greatly improved, in order  to test the theoretical predictions in the near future.

\item
Because the QCD behavior of axial-vector $a_1$ meson and that of vector $\rho$ meson
are analogous
to each other, the numerical results show the similar pattern of longitudinal
polarization between $B \to a_1\, a_1$ and $B \to \rho \rho$~\cite{Li06:b2rhorho,Chen06:b2rhorho} decays:
$B^0 \to a_1^\pm\, a_1^\mp$ and $B^\pm \to a_1^\pm\, a_1^0$ decays are dominated by the
longitudinal component, reaching around 76\%
and 91\%, respectively,
while $B^0 \to a_1^0\, a_1^0$ is governed by the transverse one at leading order, 
$f_L \approx 12\%$.
Maybe the situation for $B^0 \to a_1^0 a_1^0$ will be highly improved
 after taking the higher QCD corrections
into account. (See $B^0 \to \rho^0 \rho^0$~\cite{Li05:b2rhorho} for example.)

\item
Similar to $B^0 \to a_1^+ a_1^-$ and $B^+ \to a_1^+ a_1^0$ modes, $B^0 \to b_1^+ b_1^-$ and
$B^+ \to b_1^+ b_1^0$ ones are also dominated by the longitudinal polarization components.
However, dramatically different from $B^0 \to a_1^0\, a_1^0$ decay, $B^0 \to b_1^0\, b_1^0$
channel is absolutely governed by the longitudinal polarization contributions, say, $f_L \approx 100\%$.

\item
For a clear clarification to the polarization fractions,
we have listed the contributions from every topology with three polarizations
in the Tables~\ref{tab:DA-pm}-\ref{tab:DA-00}.
Generally speaking, from Tables~\ref{tab:DA-pm}-\ref{tab:DA-00}, one can find that relative to
$B \to a_1 a_1$ decays, all three $B \to b_1 b_1$ decays suffer from large nonfactorizable
spectator tree
contributions in the longitudinal component, which therefore lead to
the dominance of longitudinal polarization fraction. In contrast to $B^0 \to b_1^0 b_1^0$ decay,
$B^0 \to a_1^0 a_1^0$ channel receives a bit large contributions from nonfactorizable spectator
tree diagrams in both of transverse polarizations.

\end{enumerate}

We expect the above observations would be tested by the future experiments, then could
provide more information for understanding the underlying helicity structure
in these types of decays.

\subsection{Effects of Nonfactorizable Spectator and Annihilation Contributions}\label{sec:3-c}

To see whether weak annihilation
contributions
play important role in these considered decay
modes, we test the CP-averaged BRs and longitudinal polarization fraction by
neglecting the annihilation
diagrams, which can not be perturbatively calculated in the QCDF approach. Moreover,
as claimed in the references within the framework of QCDF, the hard spectator scattering
contributions also suffer from endpoint singularities at the level of twist-3.
In other words, the calculations with these terms in the QCDF need always the adjustments based on
the measurements at relevant experiments.

Without the annihilation contributions in both $B^0 \to a_1^+ a_1^-$
and $B^0 \to a_1^0 a_1^0$ decays,
we find the following CP-averaged BRs and longitudinal
polarization fractions through the numerical evaluations in the pQCD approach,
 \beq
{\rm BR}(B^0 \to a_1^+ a_1^-)&\approx& 51.0 \times 10^{-6}\;,
\qquad f_L(B^0 \to a_1^+ a_1^-)\approx 0.77\;; \label{eq:BR-noa-a1pa1m} \\
{\rm BR}(B^0 \to a_1^0 a_1^0)&\approx& \hspace{0.20cm}1.5 \times 10^{-6}\;,\;\
\qquad f_L(B^0 \to a_1^0 a_1^0)\approx 0.07\;.
 \eeq
which means that the annihilation contributions account for a 
small ratio
and could be neglected safely in $B^0 \to a_1^+ a_1^-$ mode, while it is
not the case in the decay of $B^0 \to a_1^0 a_1^0$.
When we neglect the decay amplitudes arising from both of nonfactorizable spectator
diagrams and annihilation ones, the numerical results for the CP-averaged BRs and
longitudinal polarization fractions of $B \to a_1 a_1$ decays are as follows:
\beq
{\rm BR}(B^0 \to a_1^+ a_1^-)&\approx& 49.6 \times 10^{-6}\;,
\qquad f_L(B^0 \to a_1^+ a_1^-)\approx 0.84\;;\\
{\rm BR}(B^+ \to a_1^+ a_1^0)&\approx& 20.1 \times 10^{-6}\;,\;
\qquad f_L(B^0 \to a_1^+ a_1^0)\approx 0.84\;;\\
{\rm BR}(B^0 \to a_1^0 a_1^0)&\approx& \hspace{0.20cm}0.5 \times 10^{-6}\;,\;\;
\qquad f_L(B^0 \to a_1^0 a_1^0)\approx 0.88 \;.\label{eq:BR-noa-nonsf-a10a10}
\eeq
One can find that the
nonfactorizable spectator contributions in the $B^0 \to a_1^+ a_1^-$ and $B^+ \to a_1^+
a_1^0$ modes are so small that they could be neglected, while that in
 $B^0 \to  a_1^0 a_1^0$ is large and play an important role.
This phenomenon can be easily found from the decay amplitudes shown in
Tables~\ref{tab:DA-pm}-\ref{tab:DA-00}.
In the above Eqs.~(\ref{eq:BR-noa-a1pa1m})-(\ref{eq:BR-noa-nonsf-a10a10}),
only the central values are quoted for clarification.

Similarly,
we predict the CP-averaged BRs and longitudinal polarization fractions without
the factorizable and nonfactorizable annihilation diagrams
 for $B^0 \to b_1^+ b_1^-, b_1^0 b_1^0$
decays,
 \beq
{\rm BR}(B^0 \to b_1^+ b_1^-)&\approx& \hspace{0.20cm}9.5 \times 10^{-6}\;,
\qquad f_L(B^0 \to b_1^+ b_1^-)\approx 0.74\;; \label{eq:BR-noa-b1pb1m} \\
{\rm BR}(B^0 \to b_1^0 b_1^0)&\approx& 18.7 \times 10^{-6}\;,\;
\qquad f_L(B^0 \to b_1^0 b_1^0)\approx 0.99\;.
 \eeq
which means that there should exist large contributions from annihilation diagrams
in these two decays.
Meanwhile, they also indicate that there are large nonfactorizable spectator diagrams~\cite{Diehl01:b2aa} due
to the fact of large longitudinal polarization fractions and extremely small or vanished decay constant in the
longitudinal twist-2 wave function. By considering only factorizable
emission diagrams in Fig.~\ref{fig:fig1}, the predicted BRs in the pQCD
approach are determined completely by the transverse components because the longitudinal contributions
from Fig.~1(a) and 1(b) are
sharply suppressed by the tiny or zero decay constant,
 \beq
{\rm BR}(B^0 \to b_1^+ b_1^-)&\approx& 2.6 \times 10^{-6}\;,\label{eq:BR-noa-nonsf-b1pb1m}
\qquad f_L(B^0 \to b_1^+ b_1^-)\approx 0.0\;;\\
{\rm BR}(B^+ \to b_1^+ b_1^0)&\approx& 1.0 \times 10^{-6}\;,\;\label{eq:BR-noa-nonsf-b1pb10}
\qquad f_L(B^0 \to b_1^+ b_1^0)\approx 0.0\;;\\
{\rm BR}(B^0 \to b_1^0 b_1^0)&\approx& 3.0 \times 10^{-8}\;,\;\;
\qquad f_L(B^0 \to b_1^0 b_1^0)\approx 0.0 \;.\label{eq:BR-noa-nonsf-b10b10}
\eeq
which exhibit evidently the dominated spectator and/or annihilation contributions
 involved in the $B \to b_1 b_1$ channels.
Notice that for $B^0 \to b_1^0 b_1^0$ the CP-averaged branching ratio
is significantly small just because the emission decay amplitudes are proportional to
the combined Wilson coefficient $a_2$ in both transverse polarizations.

\subsection{Direct CP-violating Asymmetries}\label{sec:3-d}

Now we turn to the evaluations of the CP-violating asymmetries for $B
\to a_1 a_1$ and $b_1 b_1$ decays in the pQCD approach.
It is conventional to combine the three polarization fractions in Eq.~(\ref{eq:pfs})
with those of its CP-conjugate $\bar{B}$ decay, and to quote the six resulting
observables corresponding to tranversity amplitudes as direct
induced CP asymmetries\footnote{The direct CP asymmetries in
transversity basis can be defined as
 \beq
 \acp^{\rm dir,\alpha}=
 \frac{\bar f_\alpha- f_\alpha}{\bar f_\alpha+
 f_\alpha},(\alpha=L,\parallel,\perp)
 \eeq
where the definition of
$\bar f$ is same as that in~Eq.(\ref{eq:pfs}) but for the
 corresponding $\bar B$ decay.}~\cite{Beneke07:b2vv}.

As for the direct CP-violating asymmetry in these considered modes, considering the
involved three polarizations, whose definitions are as follows,
 \beq
 \acp^{\rm dir} &\equiv& \frac{\bar\Gamma - \Gamma}{\bar\Gamma+ \Gamma}
 = \frac{|\overline{M}(\bar{B} \to \bar{f}_{\rm final})|^2 - |M(B \to f_{\rm final})|^2}
        {|\overline{M}(\bar{B} \to \bar{f}_{\rm final})|^2 + |M(B \to f_{\rm final})|^2}\;, \label{eq:dcp-total}
        \eeq
where $\Gamma$ and $M$ denote the decay rate and decay amplitude of $B \to a_1 a_1, b_1 b_1$ decays,
respectively, and $\bar \Gamma$ and $\overline{M}$ are the charge conjugation one correspondingly.

Based on the above definitions on direct CP-violating asymmetry and numerical
calculations in the pQCD approach(see Tables~\ref{tab:a1pa1m}-\ref{tab:a10a10}),
some remarks are as follows:

\begin{enumerate}

\item
The direct CP asymmetries for $B \to a_1 a_1$ decays in the pQCD approach can be read as,
   \beq
\acp^{\rm dir} (B^0 \to a_1^\pm a_1^\mp)
&\approx& -0.04 \pm 0.01 \;, \\
\acp^{\rm dir} (B^\pm \to a_1^\pm a_1^0) &\approx& 0.00\;,\\
\acp^{\rm dir} (B^0 \to a_1^0 a_1^0) &\approx& -0.78^{+0.08}_{-0.04}\;.
 \eeq
which are very similar to those in the $B \to \rho \rho$ decays~\cite{Li06:b2rhorho} correspondingly, where
the various errors as specified have been added in quadrature.

\item
As for the direct CP-violating asymmetries of $B \to b_1\, b_1$ decays,
   \beq
\acp^{\rm dir} (B^0 \to b_1^\pm b_1^\mp)
&\approx& -0.00^{+0.01}_{-0.04} \;, \\
\acp^{\rm dir} (B^\pm \to b_1^\pm b_1^0) &\approx& 0.00\;,\\
\acp^{\rm dir} (B^0 \to b_1^0 b_1^0) &\approx& -0.03^{+0.03}_{-0.01}\;.
 \eeq
One can find
that the numerical results in the pQCD
approach at leading order
are very small, even to be zero within uncertainties as presented
in Tables~\ref{tab:a1pa1m}-\ref{tab:a10a10}.

\item
Meanwhile, we calculate the direct CP-violating asymmetries in every polarization and
give the results in the pQCD approach as
\beq
\acp^{{\rm dir}, L} &=& 0.11^{+0.02}_{-0.02}\;, \qquad \acp^{{\rm dir}, ||} = -0.52^{+0.13}_{-0.10}\;, \qquad
\acp^{{\rm dir}, \perp} = -0.54^{+0.14}_{-0.09}\;;
\eeq
for $B^0 \to a_1^\pm a_1^\mp$ mode, and
\beq
\acp^{{\rm dir}, L} &=& 0.20^{+0.10}_{-0.55}\;, \qquad \acp^{{\rm dir}, ||} = -0.91^{+0.10}_{-0.04}\;, \qquad
\acp^{{\rm dir}, \perp} = -0.90^{+0.11}_{-0.04}\;;
\eeq
for $B^0 \to a_1^0 a_1^0$ channel, and
\beq
\acp^{{\rm dir}, L} &=& -0.05^{+0.04}_{-0.05}\;, \qquad \acp^{{\rm dir}, ||} = 0.34^{+0.05}_{-0.08}\;, \qquad
\acp^{{\rm dir}, \perp} = 0.38^{+0.07}_{-0.07}\;;
\eeq
for $B^0 \to b_1^\pm b_1^\mp$ mode, and
\beq
\acp^{{\rm dir}, L} &=& -0.03^{+0.03}_{-0.02}\;, \qquad \acp^{{\rm dir}, ||} = 0.19^{+0.30}_{-0.38}\;, \qquad
\acp^{{\rm dir}, \perp} = 0.23^{+0.30}_{-0.38}\;;
\eeq
for $B^0 \to b_1^0 b_1^0$ channel, respectively,
in which
the various errors as specified have also been added in
quadrature. These direct CP-violating asymmetries are expected to be confronted with
the relevant measurements in the future.

\item
Because of the lack of strong phase from the
annihilation diagrams and the rather negligible contributions just from electroweak penguin operators
in nonfactorizable spectator diagrams,
the direct CP violations in the $B^\pm \to a_1^\pm a_1^0$ and $b_1^\pm b_1^0$ decays
is absent for every polarization naturally, which can be seen easily in
Table~\ref{tab:DA-p0}.

\end{enumerate}

We also define another two quantities to reflect the existence of direct CP-violating
asymmetries indirectly,
 \beq
 \Delta\phi_{||} &=& \frac{\bar\phi_{||}- \phi_{||}}{2} \; \hspace{0.5cm}
  {\rm and} \hspace{0.5cm}
 \Delta\phi_{\perp} = \frac{\bar\phi_{\perp}- \phi_{\perp}- \pi}{2} \;, \label{eq:drps}
 \eeq
where $\bar\phi_{||}$ and $\bar\phi_{\perp}$ are the CP-conjugated ones of relative phases
$\phi_{||}$ and $\phi_{\perp}$, respectively. Based on the definitions of transversity amplitudes, the relative phases $\phi_{||}$ and
$\phi_{\perp}$ are defined as,
  \beq
  \phi_{||} &\equiv& \arg{\frac{{\cal A}_{||}}{{\cal A}_L}}\; \hspace{0.5cm}
  {\rm and} \hspace{0.5cm}
  \phi_{\perp} \equiv \arg{\frac{{\cal A}_{\perp}}{{\cal A}_L}}\;, \label{eq:rps}
  \eeq
The theoretical predictions of relative phases for $B \to a_1\, a_1$ and $b_1\, b_1$ 
modes in
the pQCD approach have been presented in Tables~~\ref{tab:a1pa1m}-\ref{tab:a10a10},
which will be tested by the measurements at B factories, ongoing LHC, even forthcoming
Super-B experiments.
Note that the definitions of ${\cal A}_{L,\parallel,\perp}$ as given in
Eq.~(\ref{eq:ase}) are consistent with those in~\cite{Cheng08:b2aa},
except for an additional minus sign in ${\cal A}_{L}$, so that our
definitions of the relative strong phases $\phi_{\parallel,\perp}$
(see Tables~\ref{tab:a1pa1m}-\ref{tab:a10a10}) also differ from the ones
in~\cite{Cheng08:b2aa} by $\pi$, which is added to cancel the additional
minus sign in the definition of ${\cal A}_L$ in Eq.~(\ref{eq:ase}).
\vspace{0.5cm}

At last,  it is worth of stressing that the theoretical predictions in
the pQCD approach still have large theoretical errors(See Table~\ref{tab:a1pa1m} for example)
mainly induced by the still large uncertainties of distribution amplitudes from
the shape parameter $\omega_B$ of heavy $B$ meson and
the Gegenbauer moments $a_i^{||(\perp)}$ of light axial-vector $a_1$ and $b_1$ mesons.
We need the nonperturbative QCD efforts and experimental constraints
to effectively reduce the errors of these essential inputs.
Any progress at this aspect will help us to improve the precision of the pQCD predictions.


\section{Summary}\label{sec:sum}

In this work, we studied the charmless hadronic $B \to
a_1 a_1$ and $b_1 b_1$ decays by employing the pQCD
approach based
on the $k_T$ factorization theorem. We calculated not only the factorizable
emission diagrams, but also the nonfactorizable spectator and annihilation ones.
Our theoretical predictions in the pQCD approach will provide an important platform for
testing the SM and exploring the helicity structure of these considered
decays and the hadronic dynamics of the axial-vector $a_1$ and $b_1$ mesons.
They can also provide more information on measuring the unitary CKM angles
and understanding
the decay mechanism of color-suppressed modes.

The pQCD predictions for $B \to a_1 a_1,\, b_1 b_1$ channels are displayed in
Tables~(\ref{tab:a1pa1m}-\ref{tab:a10a10}). From our
evaluations and phenomenological analysis, we found the following results:
\begin{itemize}

\item
The CP-averaged branching ratio of $B^0 \to a_1^+ a_1^-$ mode in the pQCD approach
 is in good consistency with that given by preliminary measurement
 and that presented in the QCDF framework, respectively, within errors.

\item
The pQCD predictions for the CP-averaged branching ratios of $B \to a_1 a_1,\, b_1 b_1$
decays are in the
range of $10^{-5}$ to $10^{-6}$, which can be easily accessed
at the B factories of BaBar and Belle, running LHC, and
forthcoming Super-B experiments.

\item
The numerical results in the pQCD approach, specifically, on the CP-averaged branching ratios
and longitudinal polarization fractions of the considered
$B \to a_1 a_1,\, b_1 b_1$ decays are basically consistent
with those given in the QCDF framework, except for $f_L(B^0 \to a_1^0 a_1^0)$.

\item
The theoretical predictions in the pQCD approach have large uncertainties, which
mainly arise from the nonperturbative input parameters with still large errors,
for example, the distribution amplitudes describing
the hadron dynamics of the involved mesons.
We expect these inputs will be well constrained when more data become available.

\item
We here simply take the short-distance contributions into account
in the evaluations of $B \to a_1 a_1,\, b_1 b_1$ decays.
Maybe the final state interactions for these considered modes play an important role,
more relevant studies are therefore helpful for us to provide reliable pQCD predictions.

\end{itemize}

\begin{acknowledgments}

The authors would like to thank Professor Hai-Yang~Cheng for helpful discussions.
X.~Liu is grateful to W.~Wang and Y.M.~Wang for their comments.
This work is supported by the National Natural Science
Foundation of China under Grant No.~11205072, No.~10975074, and No.~11235005, by a project funded by the
Priority Academic Program Development
of Jiangsu Higher Education Institutions (PAPD),
and by the Research Fund of Jiangsu Normal University under Grant No.~11XLR38.

\end{acknowledgments}


\end{document}